\documentclass{aa}  

\usepackage[T1]{fontenc}
\usepackage{graphicx}
\usepackage{natbib}

\newcommand{\teff}{\mbox{$T_{\rm eff}$}}
\newcommand{\logg}{\mbox{$\log g$}}

\newcommand{\kms}{\mbox{km\,s$^{-1}$}}

\newcommand{\degree}{\ensuremath{^\circ}}

\def\ms{\hbox{\,m\,s$^{-1}$}}         
\def\m2s2{\hbox{\,m$^{2}$\,s$^{-2}$}} 
\def\kms{\hbox{\,km\,s$^{-1}$}}       
\def\Msun{\hbox{$M_{\odot}$}}             

\def \1s{$1\,\sigma$}

\def \t0{T$_0$}

\newcommand{\mearth}{{\hbox{$M_{\oplus}$}}}
\newcommand{\rearth}{{\hbox{$R_{\oplus}$}}}

\usepackage{txfonts}
\begin{document} 

   \title{Precise masses for the transiting planetary system HD~106315 with HARPS \thanks{Based on observations collected at the European Organisation for Astronomical Research in the Southern Hemisphere 
under ESO programme 198.C-0168.} }
   \author{  S.~C.~C.~Barros\inst{\ref{IA}}\thanks{E-mail: susana.barros@astro.up.pt.}
          \and H.~Gosselin\inst{\ref{LAM}}\fnmsep\inst{\ref{Toulouse}}
          \and J.~Lillo-Box\inst{\ref{ESO}}
          \and D.~Bayliss\inst{\ref{geneva}}
          \and E.~Delgado~Mena\inst{\ref{IA}}
         \and B.~Brugger\inst{\ref{LAM}} 
         \and A.~Santerne\inst{\ref{LAM}}        
          \and D.~J.~Armstrong\inst{\ref{warwick}}
          \and V.~Adibekyan\inst{\ref{IA}}       
        \and J.~D.~Armstrong\inst{\ref{hawai}}
          \and D.~Barrado\inst{\ref{CAB}}
          \and J.~Bento\inst{\ref{australia}}
          \and I.~Boisse\inst{\ref{LAM}}
          \and A.~S.~Bonomo\inst{\ref{Torino}}
          \and F.~Bouchy\inst{\ref{geneva}}
          \and D.~J.~A.~Brown\inst{\ref{warwick}}
         \and  W. D. Cochran\inst{\ref{Texas}}
         \and A.~Collier Cameron\inst{\ref{standrews}}
          \and M.~Deleuil\inst{\ref{LAM}}        
          \and O.~Demangeon\inst{\ref{IA}}
          \and R.~F.~D\'iaz\inst{\ref{geneva}}\fnmsep\inst{\ref{argentina}}\fnmsep\inst{\ref{CONICET}}
          \and A.~Doyle\inst{\ref{warwick}}
          \and X.~Dumusque\inst{\ref{geneva}}
         \and D.~Ehrenreich\inst{\ref{geneva}}
         \and  N.~Espinoza\inst{\ref{chile}}\fnmsep\inst{\ref{chile2}}
          \and F.~Faedi\inst{\ref{warwick}}
          \and J.~P.~Faria\inst{\ref{IA}}\fnmsep\inst{\ref{UPorto}}
          \and P.~Figueira\inst{\ref{IA}}
          \and E.~Foxell\inst{\ref{warwick}}
          \and G.~H\'ebrard\inst{\ref{IAP}}\fnmsep\inst{\ref{OHP}}
          \and S.~Hojjatpanah\inst{\ref{IA}}\fnmsep\inst{\ref{UPorto}}          
          \and J.~Jackman\inst{\ref{warwick}}
         \and  M.~Lendl\inst{\ref{austria}}
          \and R.~Ligi\inst{\ref{LAM}}
          \and C.~Lovis\inst{\ref{geneva}}
        \and C.~Melo\inst{\ref{ESO}} 
          \and O.~Mousis\inst{\ref{LAM}}
          \and J.~J.~Neal\inst{\ref{IA}}\fnmsep\inst{\ref{UPorto}}
          \and H.~P.~Osborn\inst{\ref{warwick}}
          \and D.~Pollacco\inst{\ref{warwick}} 
          \and N.~C.~Santos\inst{\ref{IA}}\fnmsep\inst{\ref{UPorto}}
          \and R.~Sefako\inst{\ref{africa}}
          \and A.~Shporer \inst{\ref{caltech}}
          \and S.~G.~Sousa\inst{\ref{IA}}
          \and A.~H.~M.~J.~Triaud \inst{\ref{Cambridge}} 
          \and S.~Udry\inst{\ref{geneva}}
          \and A.~Vigan\inst{\ref{LAM}}
          \and A.~Wyttenbach\inst{\ref{geneva}}     
         }

          \institute{ Instituto de Astrof\'isica e Ci\^encias do Espa\c{c}o, Universidade do Porto, CAUP, Rua das Estrelas, PT4150-762 Porto, Portugal \label{IA} 
        \email{susana.barros@astro.up.pt}
              \and
              Aix Marseille Univ, CNRS, LAM, Laboratoire d'Astrophysique de Marseille, Marseille, France\label{LAM}
              \and
              European Southern Observatory (ESO), Alonso de Cordova 3107, Vitacura, Casilla 19001, Santiago de Chile, Chile\label{ESO}
              \and
              Observatoire Astronomique de l'Universite de Geneve, 51 Chemin des Maillettes, 1290 Versoix, Switzerland\label{geneva}
              \and
              Department of Physics, University of Warwick, Gibbet Hill Road, Coventry, CV4 7AL, UK\label{warwick}
                \and
             Institute for Astronomy, University of Hawaii, 34 Ohia Ku Street, Pukalani, Maui, Hawaii 96790\label{hawai}
              \and   
              Depto. de Astrof\'isica, Centro de Astrobiolog\'ia (CSIC-INTA), ESAC campus 28692 Villanueva de la Ca\~nada (Madrid), Spain\label{CAB}
            \and
             Research School of Astronomy and Astrophysics, Australian National University, Mount Stromlo Observatory, Cotter Road, Weston Creek, ACT 2611, Australia\label{australia}          
              \and
              INAF -- Osservatorio Astrofisico di Torino, Strada Osservatorio 20, I-10025, Pino Torinese (TO), Italy\label{Torino}
              \and
             McDonald Observatory and Department of Astronomy, The University of Texas at Austin, Austin Texas USA \label{Texas}     
               \and
             Centre for Exoplanet Science, SUPA School of Physics \& Astronomy, University of St Andrews, North Haugh ST ANDREWS, Fife, KY16 9SS  \label{standrews}
               \and
             Universidad de Buenos Aires, Facultad de Ciencias Exactas y Naturales. Buenos Aires, Argentina\label{argentina}
             \and
             CONICET - Universidad de Buenos Aires. Instituto de Astronom\'ia y F\'isica del Espacio (IAFE). Buenos Aires, Argentina.\label{CONICET}     
              \and 
              Instituto de Astrofisica, Facultad de Fisica, Pontificia Universidad Catolica de Chile, Av. Vicuna Mackenna 4860, 782-0436 Macul, Santiago, Chile \label{chile}
              \and
             Millennium Institute of Astrophysics, Av. Vicuna Mackenna 4860, 782-0436 Macul, Santiago, Chile \label{chile2}                   
              \and
              Departamento\,de\,Fisica\,e\,Astronomia,\,Faculdade\,de\,Ciencias,\,Universidade\,do\,Porto,\,Rua\,Campo\,Alegre,\,4169-007\,Porto,\,Portugal \label{UPorto}
              \and
              Institut d'Astrophysique de Paris, UMR7095 CNRS, Universite Pierre \& Marie Curie, 98bis boulevard Arago, 75014 Paris, France\label{IAP}
              \and
              Aix Marseille Univ, CNRS, OHP, Observatoire de Haute Provence, Saint Michel l'Observatoire, France\label{OHP}
              \and
              Space Research Institute, Austrian Academy of Sciences, Schmiedlstr. 6, 8042, Graz, Austria\label{austria}
               \and
             South African Astronomical Observatory, PO Box 9, Observatory, 7935\label{africa}
               \and
              Division of Geological and Planetary Sciences, California Institute of Technology, Pasadena, CA 91125, USA\label{caltech}
              \and 
              Institute of Astronomy, University of Cambridge, Madingley Road, CB3 0HA, Cambridge, United Kingdom\label{Cambridge}
              \and
             Universit\'e de Toulouse, UPS-OMP, IRAP, Toulouse, France\label{Toulouse}           
             }

   \date{Received ??, ??; accepted ??}
  \abstract
   {  The multi-planetary system HD~106315 was recently found in K2 data . The planets have periods of $P_b \sim9.55$ and $P_c \sim 21.06\,$days, and radii of $ r_b = 2.44  \pm 0.17\, $ \rearth\ and  $r_c = 4.35 \pm 0.23\, $ \rearth. The brightness of the host star (V=9.0 mag) makes it an excellent target for transmission spectroscopy. However, to interpret transmission spectra it is crucial to measure the planetary masses.}
   { We obtained high precision radial velocities for HD~106315 to determine the mass of the two transiting planets discovered with Kepler K2.  Our successful observation strategy was carefully tailored to mitigate the effect of stellar variability. }    
   { We modelled the new radial velocity data together with the K2 transit photometry and a new ground-based partial transit of HD~106315c to derive system parameters.}
   {We estimate the mass of HD~106315b to be 12.6 $\pm$ 3.2 \mearth\ and the density to be $4.7 \pm1.7\, g\,cm^{-3}$, while for HD~106315c we estimate a mass of  15.2 $\pm$ 3.7 \mearth\ and  a density of $1.01 \pm 0.29\, $g\,cm$^{-3}$.  Hence, despite planet c having a radius almost twice as large as planet b, their masses are consistent with one another.}
   {We conclude that HD~106315c has a thick hydrogen-helium gaseous envelope. A detailed investigation of HD~106315b using a planetary interior model constrains the core mass fraction to be 5-29\%, and the water mass fraction to be 10-50\%. An alternative, not considered by our model,  is that HD~106315b is composed of a large rocky core with a thick H-He envelope.  Transmission spectroscopy of these planets will give insight into their atmospheric compositions and also help constrain their core compositions.  }  
 \maketitle
%
\section{Introduction}
\label{intro}
The field of exoplanets has been revolutionised by the discovery of $\sim5000$ planetary candidates by the  {\it Kepler}  space mission \citep{Borucki2010}.
 {\it Kepler}  revealed the existence of a large diversity of exoplanets and enabled the discovery of planets even smaller than the Earth  \cite[e.g.][]{Barclay2013, Jontof-Hutter2015}.
Surprisingly, it also showed that the most common type of planets with periods less than 100 days have sizes of 1.5-4 \rearth\ (between those of the Earth and Neptune) \citep{Fressin2013}, which do not exist in the solar system.  In turn, radial velocity surveys had already found that planets with low masses were more common \citep{Mayor2011}.

Recently, much interest has been devoted to gaining insight into the composition of these planets. Formation and compositions theories predict that planets are made of four main components: H-He, ices, silicates, and iron-nickel  \cite[e.g.][]{Seager2007}.  Different combinations of these materials result in a wide range of possible radii for a given planetary mass. While planets larger than Neptune are expected to be mainly gaseous and planets equal to or smaller than the Earth are expected to be rocky, planets with sizes of  1.5-4 \rearth\  can have compositions ranging from gaseous mini Neptunes, to water worlds, to rocky super-Earths  \citep{Rogers2011, Leger2004, Valencia2006,Seager2007,Rogers2015}.  Due to the faintness of the host stars in the  {\it Kepler}  field, it was only possible to derive accurate masses for relatively few of the  {\it Kepler} candidates in the small radii regime.   
For the brightest stars it has been possible to derive mass and radius with sufficient accuracy to reveal a large diversity of planetary compositions  \citep{Carter2012, Marcy2014, Barros2014, Haywood2014, Dressing2015b, Malavolta2017}. Interestingly, planets with very similar radii have very diverse densities like, for example, Kepler-11f  ($\rho =  0.7\pm 0.4 $ g cm$^{-3}$, $r_p =  2.61 \pm0.025$, \citealt{Lissauer2011a} ) and Kepler-10c  ( $\rho = 7.1 \pm 1.0 $ g cm$^{-3}$ ,  $r_p  = 2.35 \pm0.05 $ ,\citealt{Dumusque2014}). To  better understand the diversity of planetary composition in this size regime, a larger sample of exoplanets  with well-constrained mass and radius is needed. 
A better insight into planetary composition will also make it possible to constrain planetary formation processes, since the two alternative planetary formation theories, namely core accretion \citep{Lissauer1993} and gravitational instability models \citep{Nayakshin2017}, predict different metal-gas ratio compositions.

Moreover, a statistically significant sample of well-characterised low-mass planets will help to better constrain the transition between rocky and gaseous planets. The determination of the transition between rocky and gaseous planets is a way to constrain the formation theories of small planets and has implications for habitability \citep{Alibert2014}.
Recent studies point to the transition being at 1.6 \rearth\ \citep{Weiss2014,Rogers2015, Fulton2017}, although they rely on small number statistics and hence they have a large uncertainty.

Planets with sizes between super-Earths and Neptune ( 1.5-4 \rearth\ ) are also very interesting targets for transmission spectroscopy. Fortunately, it will be possible to explore the atmospheres of low-mass planets with  next generation instruments like the James Webb Space Telescope (JWST) and  the Extremely Large Telescope (ELT), and the proposed exoplanet characterisation space missions like Fast Infrared Exoplanet Spectroscopy Survey Explorer (FINESSE) and Atmospheric Remote-sensing Exoplanet Large-survey (Ariel). The interpretation of atmospheric measurements requires  precise determination of the planetary surface gravity \citep{Batalha2017}. Therefore, it requires the measurement of both precise radius and mass for the planets.

For all these reasons,  we started the follow-up of K2 \citep{Howell2014} low-mass planetary candidates with the High Accuracy Radial velocity Planet Searcher (HARPS) spectrograph \citep{Mayor2003}. Due to its large combined field of view, K2  is finding a higher number of low-mass planets around bright stars. These bright host stars allow the derivation of planetary masses with good precision and facilitate future atmospheric follow-up observations.

The multi-planetary system HD~106315 (EPIC 201437844) was observed by K2 Campaign 10. Two teams simultaneously announced the discovery of two  planets with small radii \citep{Crossfield2017,Rodriguez2017}. The inner planet, HD~106315b, has a period of 9.55 days and a radius of $\sim 2.4$ \rearth,\ while HD~106315c, the outer planet,  has a period of 21.05 days and a radius of $ \sim  4.2$ \rearth\ .  
 \citet{Crossfield2017} obtained radial velocities of HD~106315, which did not constrain the mass of either planet but revealed a long term trend that suggests the presence of an additional third body in the system. The host star, HD~106315, has V=9.0 mag, which is currently the third brightest planet host star found by K2. The brightness of the host star and the size of both planets make it an interesting target since an accurate mass determination should be achievable with HARPS.

  This planetary system is also interesting because HD~106315c is one of the few small planet cases for which a measurement of the obliquity of the host star is possible with current facilities \citep{Rodriguez2017} using the Rossiter-McLaughlin (RM) effect  \citep{Rossiter1924, McLaughlin1924}.  Measured obliquities in hot Jupiters have lead to insight into hot-Jupiter migration mechanisms and the interaction between star and planet, which shape planetary systems. It was shown that hot-Jupiter host stars' obliquities depend on the tidal dissipation timescale \citep{Albrecht2012}, which in turn depends on the stellar effective temperature, planet-to-star mass ratio, and scaled orbital distance.  This supports planet-planet scattering as the migration mechanism for hot Jupiters, since it predicts the misalignment of the orbits \citep{Weidenschilling1996,Rasio1996,Ford2008}.  It is not yet known if a similar trend occurs for smaller mass planets. HD~106315 is a F5V type star with \teff $~\sim$ 6260 K, which is above the suggested transition temperature between systems with different regimes of the tidal dissipation timescale \citep{Winn2010}. This, combined with the low planet-to-star mass ratio, makes it an extremely interesting system to measure the RM effect.

In this paper we present radial velocity observations of HD~106315 taken with HARPS, which allow for the derivation of the mass of HD~106315b and HD~106315c. A state of the art interior model is used to depict the possible interiors of the planets, giving insight into their interior compositions.  We report our observations in Section~2 and present the stellar characterisation in Section~3. In Section~4 we describe our method to analyse the radial velocities together with the photometric data and present the results in Section~4. Finally we discuss the results in Section~5.


\section{Observations}
\label{Observations}
\subsection{K2 photometry}

\begin{figure}
\centering
\includegraphics[width=0.45\textwidth]{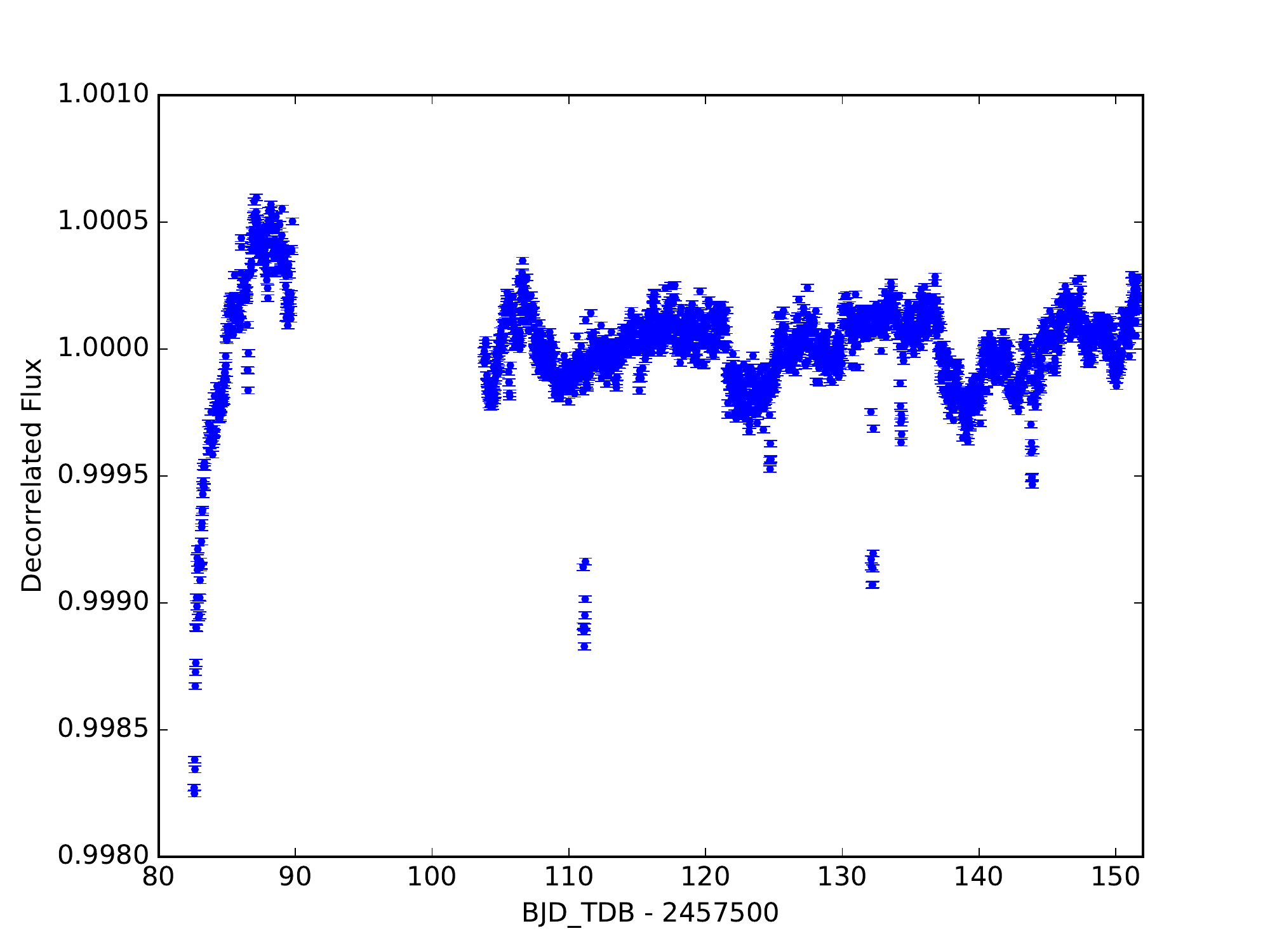}
\caption{K2 detrended light curve of HD~106315 showing short term variability and two sets of transits. \label{fig.detlc}  }  
\end{figure}

\begin{figure*}
\centering
\includegraphics[width=0.45\textwidth]{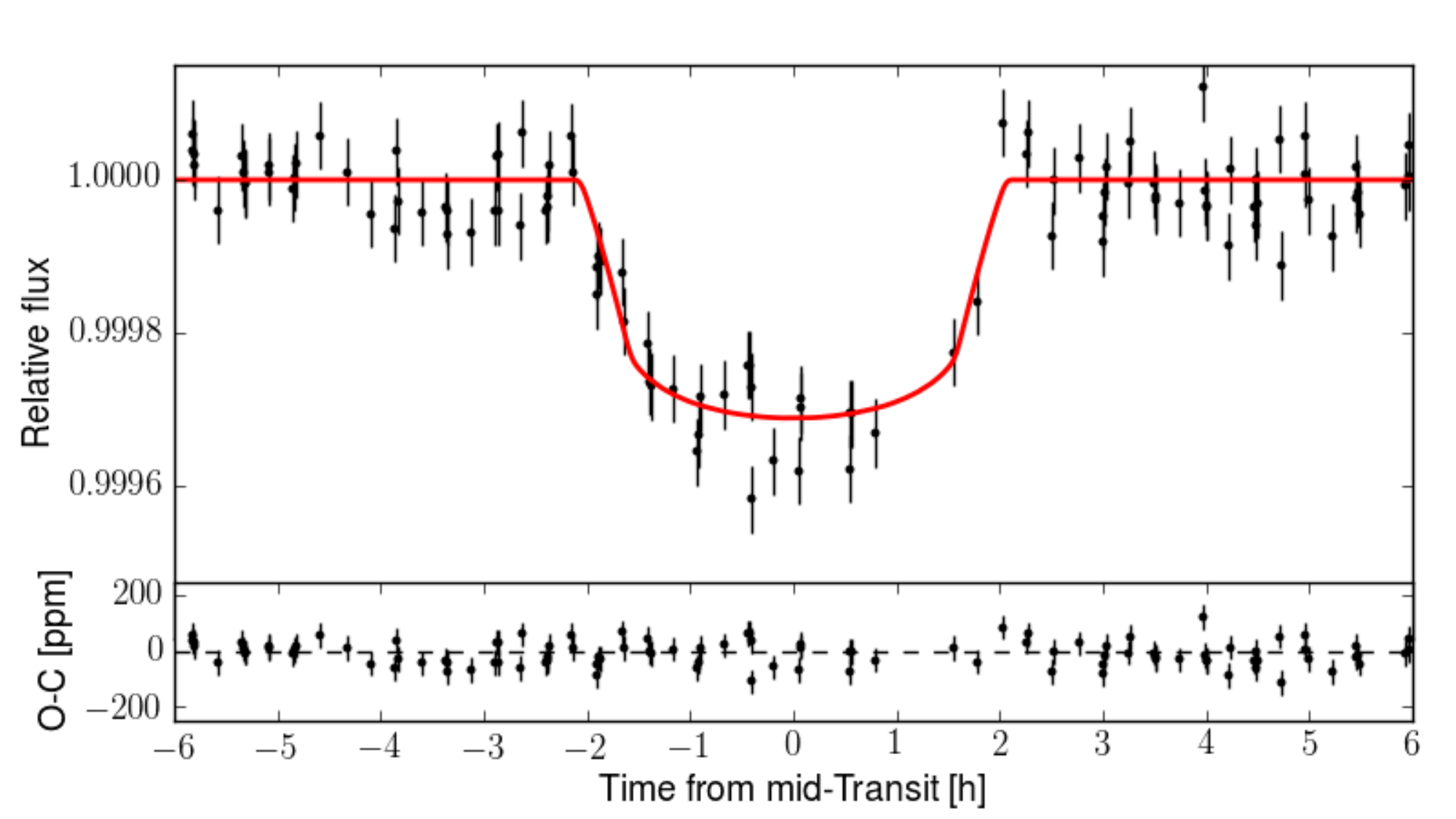}
\includegraphics[width=0.45\textwidth]{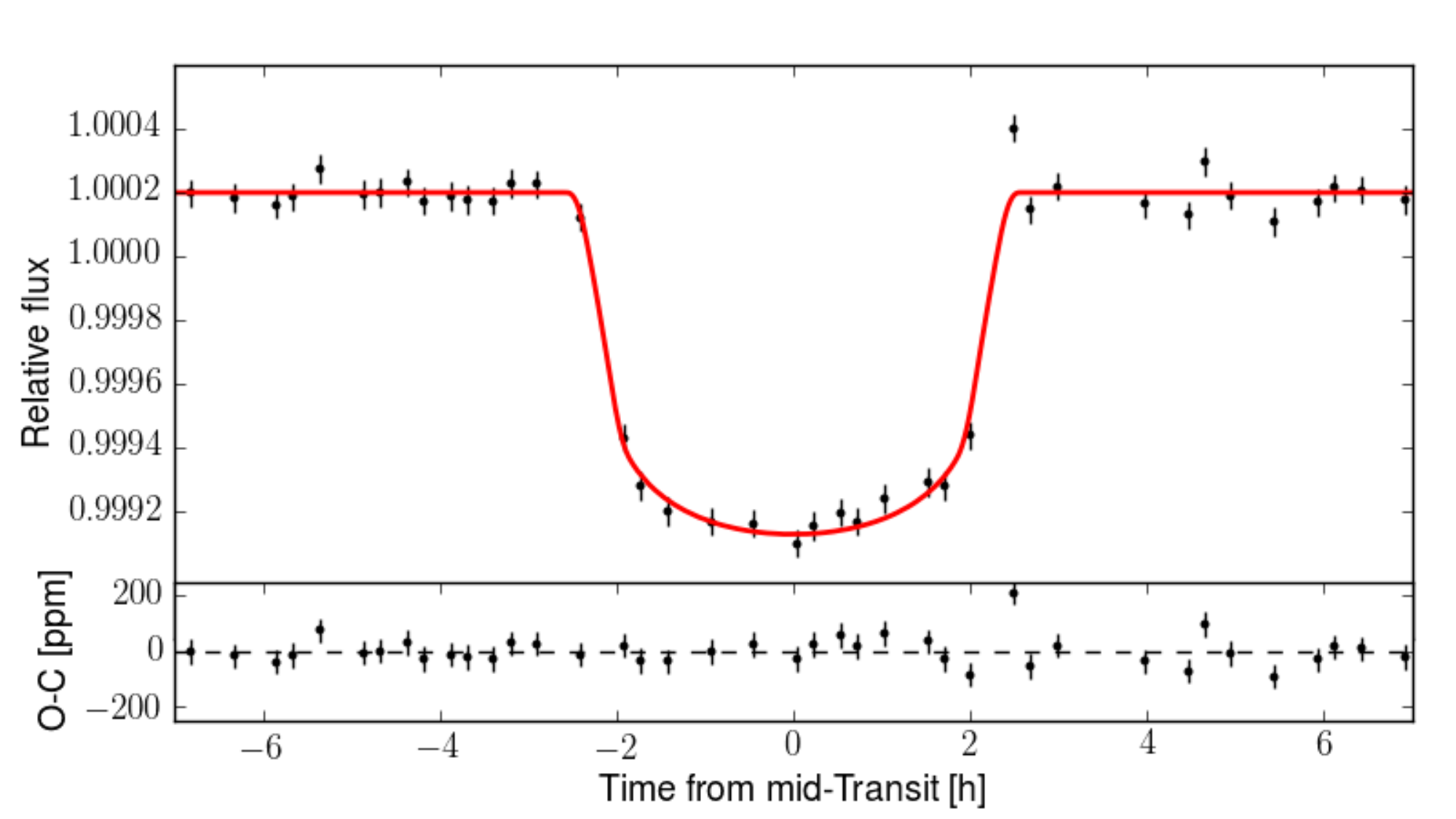}
\caption{ Phase folded K2 light curve of  HD~106315b (left panel)  and HD~106315c (right panel). We overplot the best fitted model presented in  Table~\ref{PASTISresults}   \label{transitk2}}
\end{figure*}

HD~106315 (EPIC 201437844) was observed during Campaign 10 of the K2 mission, in long cadence mode, between  2016 July 6 and 2016 September 20, spanning $\sim 80\,$ days.
However, during the first six days of Campaign 10 there was a pointing error of 3.5-pixels preventing high precision photometry, so we did not reduce this data. Furthermore, on 2016 July 20 one of the modules failed leading to the telescope going into safe mode and resulting in a 14-day gap in the observations. Therefore, Campaign 10 has slightly less data than previous campaigns.  
We downloaded the pixel data from the Mikulski Archive for Space Telescopes (MAST).\footnotemark \footnotetext{$https://archive.stsci.edu/k2/data\_search/search.php.$}
We reduced the pixel data and extracted the light curves using the Planet candidates from OptimaL Aperture Reduction (POLAR)  pipeline. The POLAR pipeline has some routines that are part of the  COnvection ROtation and planetary Transits (CoRoT) imagette pipeline \citep{Barros2014}. A full description of the POLAR  pipeline is given in \citet{Barros2016}. The POLAR reduced light curves up to Campaign 6 are publicly available through the MAST.\footnotemark \footnotetext{$https://archive.stsci.edu/prepds/polar/.$}
The K2 light curve of HD~106315 is presented in Figure~\ref{fig.detlc}. The light curve is dominated by red noise due to granulation and we cannot find a clear rotation period.
Moreover, the K2 light curve shows two sets of planetary signals that were previously reported by  \citet{Crossfield2017} and \citet{Rodriguez2017}.  Planet b shows a transit depth of $297 \pm 34$ \,ppm, while planet c shows a transit depth of $944 \pm 21$ \,ppm. The phase folded transits of K2 for both planets, together with the best transit model presented in Section~5, are shown in Figure~\ref{transitk2}. The stellar-activity-filtered final light curve has a robust rms of 55ppm.

\subsection{Follow-up photometry}

\begin{figure}
\centering
\includegraphics[width=0.45\textwidth]{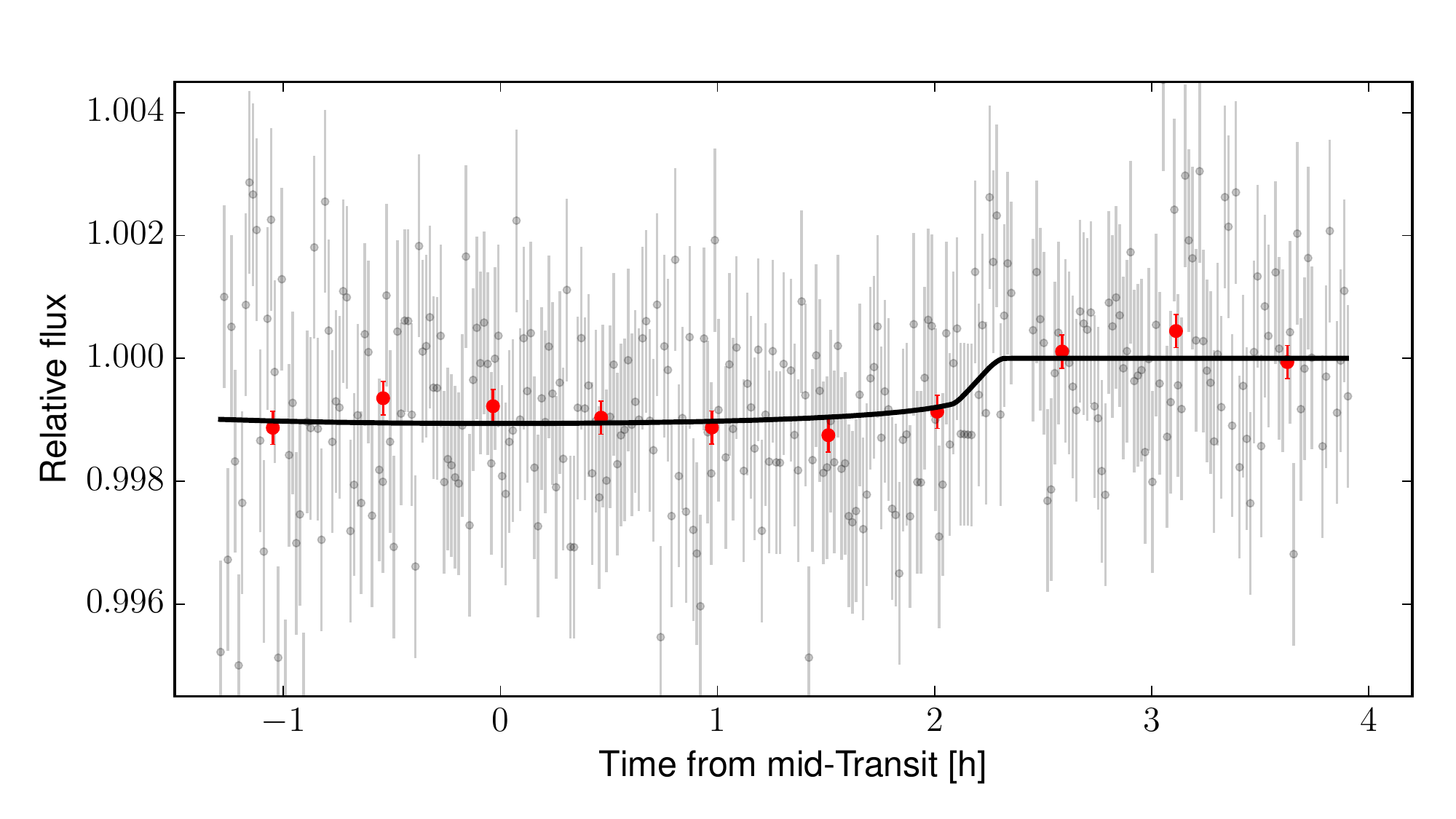}
\caption{ Phase folded LCO light curve of HD~106315c taken with the i' filter. We overplot the best fitted model presented in Table~\ref{PASTISresults}. We also overplot the binned light curve at 30 minute cadence.  The residuals of the binned light curve show an rms of 219.0 ppm. \label{transitLGTO}}
\end{figure}

In order to help constrain the orbital period and phase of HD~106315c, we monitored the transit on the night of 2017 March 8 from Cerro Tololo Inter-American Observatory (CTIO), Chile, with one of the three LCO (Las Cumbres Observatory) 1 m telescopes located at that site.  LCO operates a network of fully automated 1\,m telescopes \citep{Brown2013} equipped with custom built imaging cameras (Sinistro\footnotemark \footnotetext{$https://lco.global/observatory/instruments/sinistro/.$}) with back-illuminated 4K$\times$4K Fairchild Imaging CCDs (charged-coupled device) with $15\,\mu$m pixels (CCD486 BI).  With a plate scale of $0.387"/$pixel, the Sinistro cameras deliver a field of view of $26.6~\arcmin\times26.6~\arcmin$  , which is important for reference stars when monitoring bright stars such as HD~106315.  The cameras are read out by four amplifiers with $1\times1$ binning, with a readout time of $\approx 45$\,s.  We used the $i$-band filter, with exposure times of 30\,s.  Images were reduced by the standard LCO pipeline \citep{Brown2013}, and aperture photometry was performed in the manner set out in \cite{Penev2013}.  The phase folded light curve as well as the best fit model that we present in  Section~5  are shown in Figure~\ref{transitLGTO}. The transit shows an egress with a depth and timing consistent with the predicted ephemeris from the K2 data.

\subsection{HARPS radial velocity observations}

We observed the target star HD~106315 with the HARPS spectrograph, which is mounted on the ESO-3.6m telescope in La Silla Observatory (Chile), with the aim of detecting the mass of the two transiting planets. Since the star is relatively hot and fast-rotating, measuring precise radial velocities is challenging \citep{Rodriguez2017}. \citet{Crossfield2017} performed a first radial velocity follow-up of this target with the HiReS spectrograph at the Keck telescope. They used exposure times of three to six minutes, and observed a large rms at the level of 6.4 \ms. Hot stars are known to have large variability caused by granulation and p-modes \citep{Dumusque2011}. We estimated that p-modes have timescale of about 20 minutes, as in the case of Procyon \citep{Bouchy2004}. To average out the oscillation and granulation noises in the spectra, we used exposure times of 30 minutes and observed the target several times during the night with a few hours between exposures. 
We collected 93 spectra of HD~106315 with HARPS from 2017 January 26 to 2017 May 4,  as part of our ESO-K2 large programme (Programme ID 198.C-0169). We rejected eight spectra because they were significantly affected by the Moon background light \citep{Bonomo2010} or taken at high airmass (>1.7). We complemented our time series with 46 out-of-transit spectra taken as part of the HEARTS programme \citep{Wyttenbach2017} on HARPS (programme ID  098.C-0304) during two transits of planet c. These data have an exposure time of five minutes during the first night and eight minutes during the second night. For this dataset we excluded observations that were taken with airmass higher than 1.7. All the HARPS spectra were reduced with the HARPS pipeline consistently. Radial velocities were derived by cross-correlating the spectra with a template of a G2V star \citep{Baranne1996, Pepe2002}. Radial velocity photon noise was estimated based on the work of \cite{Bouchy2001}. We also measured the full width half maximum (FWHM) and the bisector inverse slope (BIS), and estimated their uncertainty as suggested in \citet{Santerne2015}. All the radial velocity products are reported in Table~\ref{rvdata}. To mitigate the effect of stellar variability in the HEARTs data, for the final analysis we binned them to the same exposure time as the rest of the HARPS data (30 minutes), resulting in four points in the first night after the transit and four points in the second night before the transit.

We fitted the cross-correlation function with a rotation profile as described in \citet{Santerne2012b} and derived a stellar $v \sin i = 12.71 \pm 0.4$~km~s$^{-1}$. This, combined with the stellar radius given in Table~\ref{PASTISresults}, gives a rotation period of $5.15 \pm0.28$d (if i=90).  Using the S$_{\rm MW}$ values and the calibration as described by  \citet{Noyes1984} and \citet{Lovis2011},  we derived an activity index of logR'$_{\rm HK}$ = - $4.888 \pm 0.008$.  Given that the distance to HD~106315 is higher than 100pc (Section \ref{analysis}), we checked whether interstellar clouds could be artificially decreasing the measured logR'$_{\rm HK}$. In our final fit we find a value for the extinction  of $0.0049^ {+ 0.008}_ {-0.004}$ mag; this value was confirmed by using the galaxy maps by \citet{Green2015}, which estimate the extinction is lower than 0.01. Therefore, we do not expect a bias of the  logR'$_{\rm HK}$ given the low extinction.

\begin{figure*}
\centering
\includegraphics[width=0.90\textwidth]{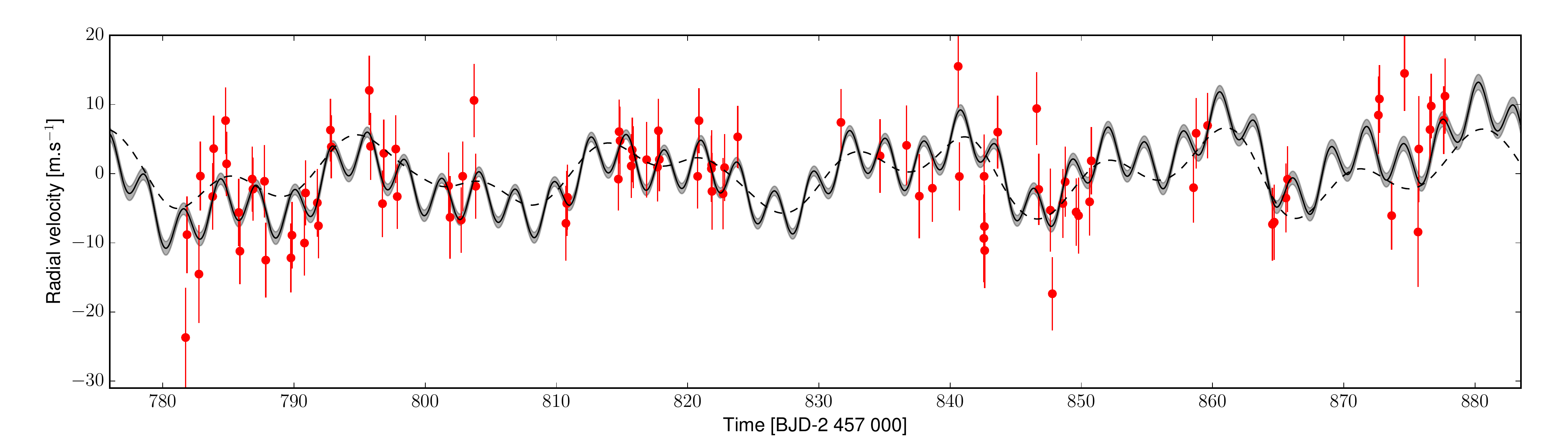}\\
\includegraphics[width=0.45\textwidth]{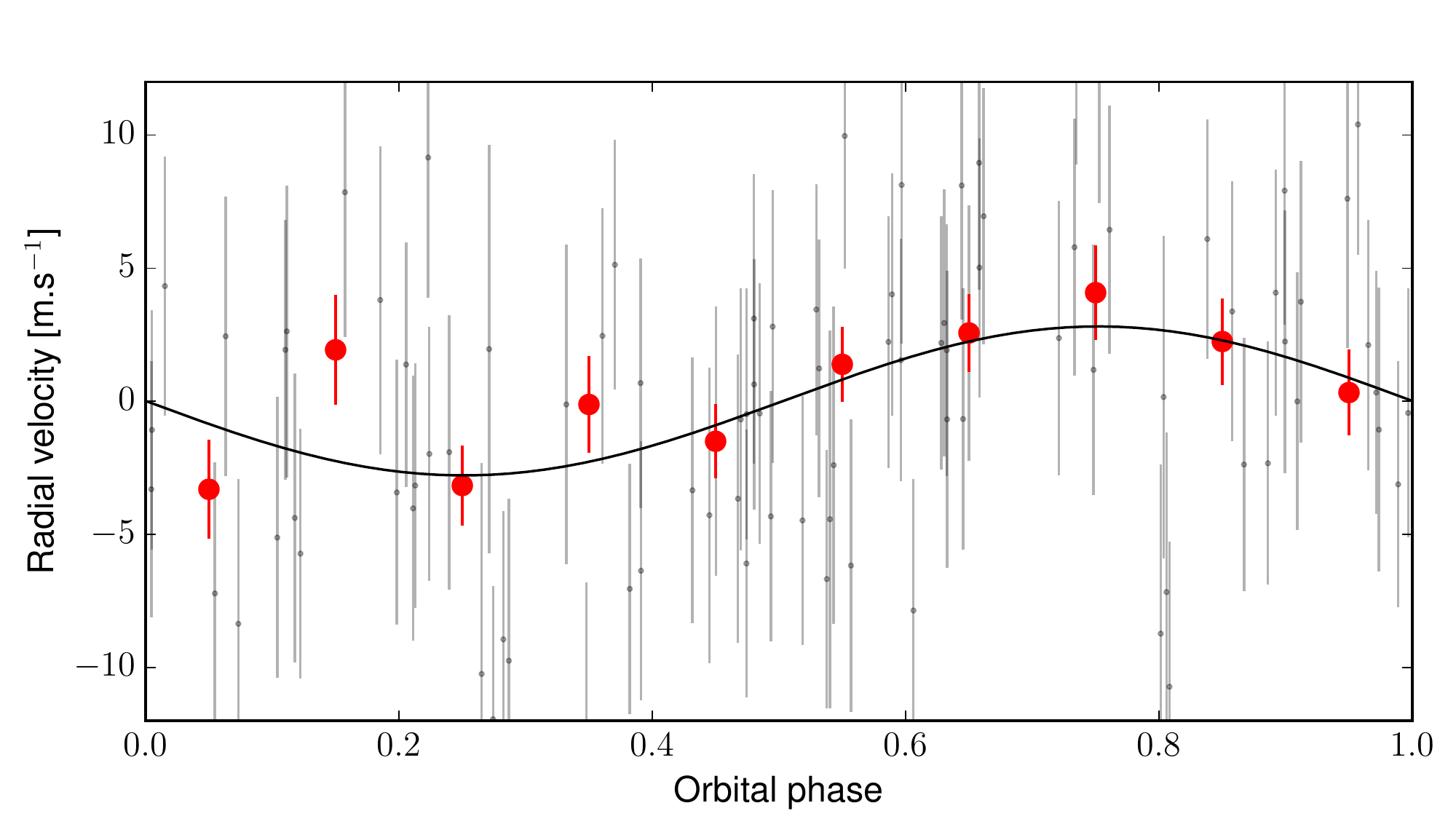}
\includegraphics[width=0.45\textwidth]{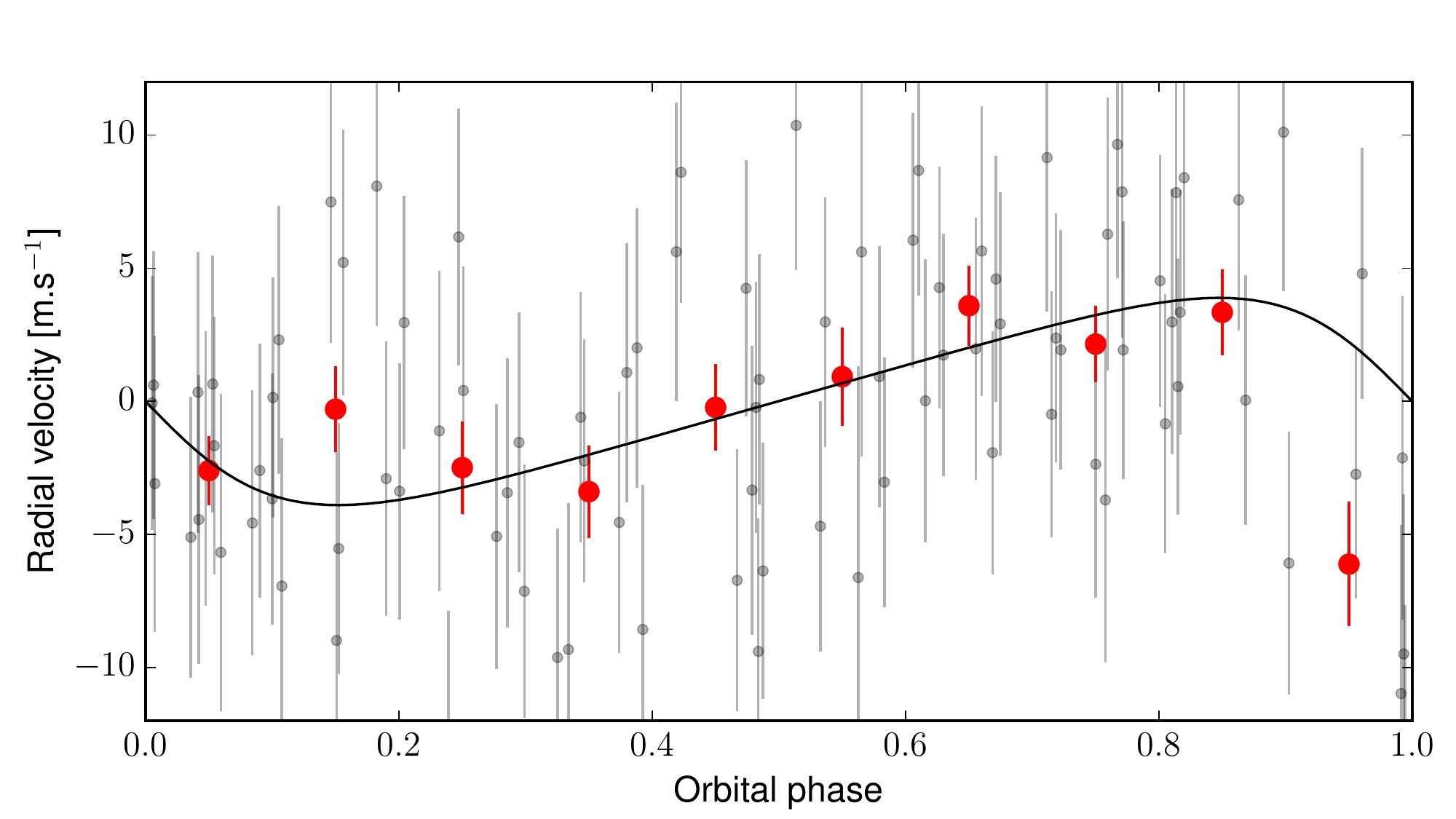}
\caption{ Top panel: Time series of the radial velocities of HD~106315. Bottom panel: Radial velocities of HD~106315 phase folded on the ephemeris of planet b (left panel)  and of planet c (right panel). For the phase folded plots, we show both data binned to 0.05 in phase (in red-dark) and unbinned data (grey).  For all the cases, the errors include the estimated jitter added quadratically and we overplotted the best fitted model presented in Table~\ref{PASTISresults},  which includes the Gaussian process to model the activity. In the top panel we also show a model without GPs, which includes the two Keplerian orbits (dash-line). \label{rvplot}}
\end{figure*}

\begin{figure}
\centering
\includegraphics[width=0.45\textwidth]{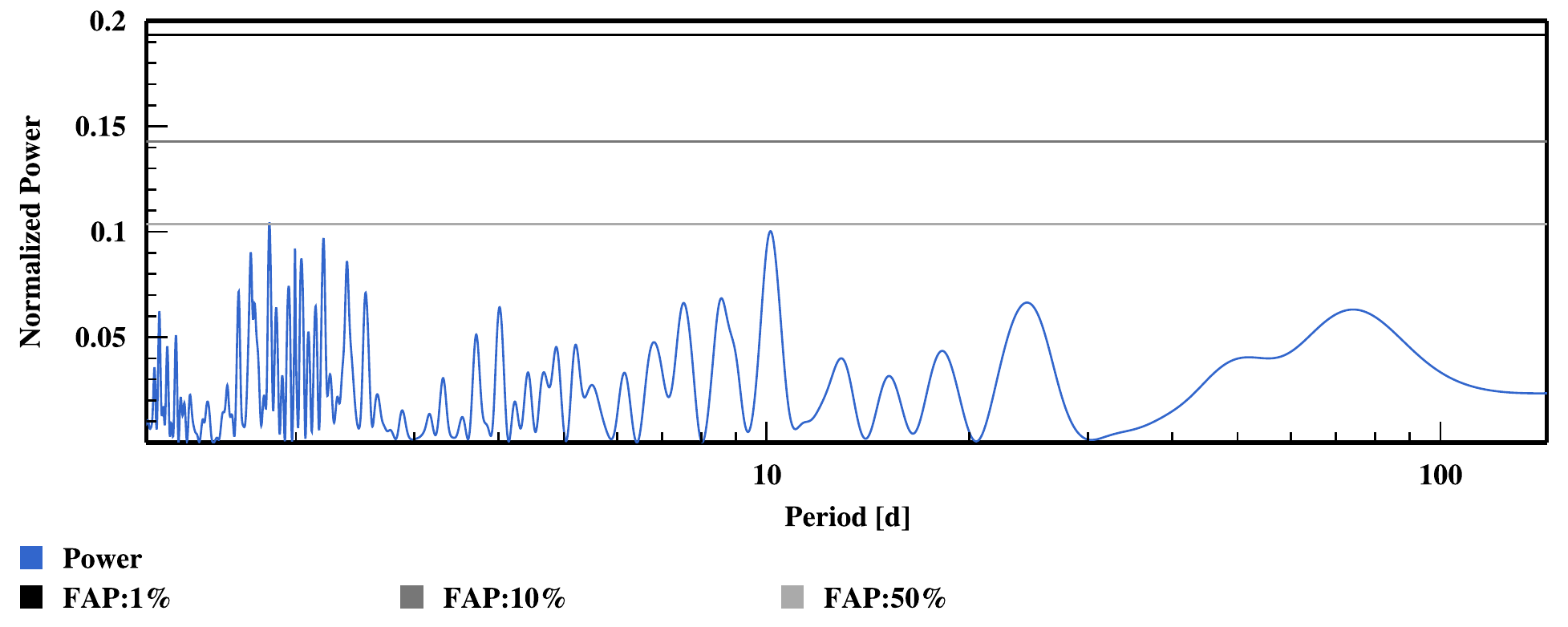}
\caption{Periodogram of the residual of the radial velocities of HD~106315 after subtracting the best fit keplerian for planets b and c. The horizontal lines represent the false alarm probability at  0.1\%, 1\%, 10\%, and 50\% confidence level. \label{fig.period}  }  
\end{figure}

The radial velocity observations show a clear variation phased with the fitted periods of the planets from the K2 photometry. The HARPS RVs together with the best fit Keplerian model (section~5 ) are shown in Figure~\ref{rvplot}.  In Figure~\ref{fig.period}, we show the periodogram of the residual of the radial velocities after subtracting the best fit model. We find no  other significant peaks in the periodogram, and hence no other planet in the system is detected.

\subsection{Lucky imaging}
\begin{figure}
\centering
\includegraphics[width=0.45\textwidth]{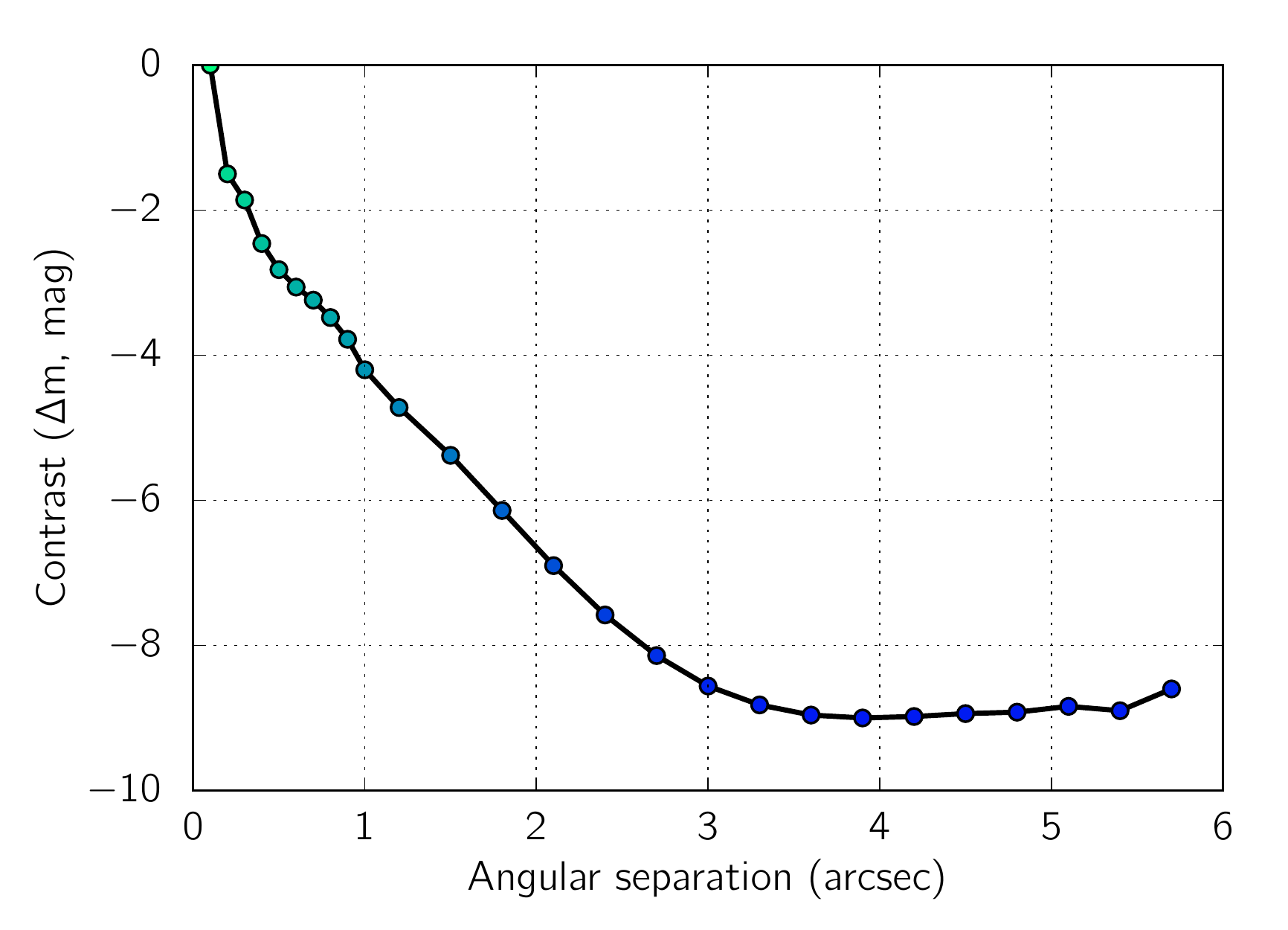}
\caption{Contrast curve in the SDSS i-band for the neighbourhood of  HD~106315.    \label{fig.astrasen}  }  
\end{figure}

We observed HD~106315 with the high spatial resolution camera AstraLux \citep{hormuth08} installed at the 2.2m telescope of Calar Alto Observatory (Almar\'ia, Spain). This instrument applies the lucky-imaging technique \citep{fried78} to avoid atmospheric distortions by obtaining thousands of short-exposure (below the atmospheric coherence time scale) images and selecting the best-quality ones based on the Strehl-ratio \citep{strehl1902}. In this case, we obtained 70\,000 images of 30\,ms exposure time in the SDSS (sloan digital sky survey) i-band and selected the best 10\% of them. We used the observatory pipeline to perform the basic reduction of the images, measure their Strehl ratio, select the best-quality frames, and finally align them and combine them to obtain the final high spatial resolution image.  In this final image, we find no additional companions up to 12 arcsec within the sensitivity limits. In Figure~\ref{fig.astrasen} we show the sensitivity curve determined following the process explained in \citet{lillo-box14b}, based on the injection of artificial stars into the image at different angular separations and position angles, and measuring the retrieved stars based on the same detection algorithms used to look for real companions.

\begin{figure}
\centering
\includegraphics[width=0.45\textwidth]{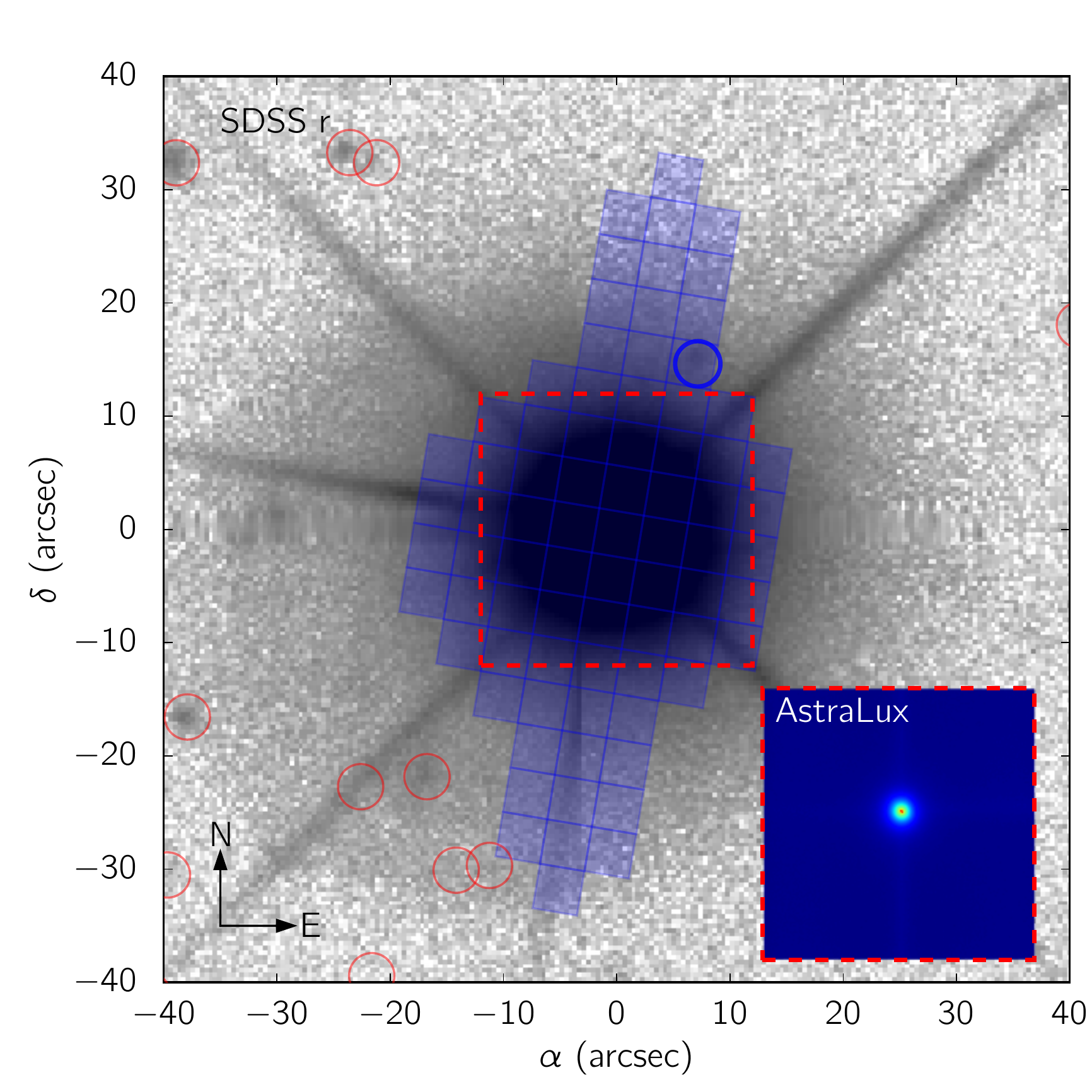}
\caption{ SDSS r' band image of the field of HD~106315 with an inset showing the high spacial resolution image taken with AstraLux. Only one faint object (marked with a blue circle) is detected inside the K2 aperture (shadowed purple). The object is 11 magnitudes fainter than the main target and hence does not produce significant contamination.  \label{fig.astraimg} }
\end{figure}

Due to HD~106315 being saturated in K2, the best aperture used to extract the photometry is quite large, with a maximum eight pixels in the x direction and 17 pixels in the y direction that correspond to 32x68 arcsec. Since this aperture is larger than the field of view of AstraLux (24x24  arcsec), we also checked for companions that could be inside the K2 aperture using archival images. In Figure~\ref{fig.astraimg} we overlay the K2 aperture, the AstraLux image, and the SDSS r' band image. Detected sources are marked with red circles in the image. We find a star at the edge of the K2 aperture (marked with a blue circle) that is 11 mag fainter than the target in the r' band. Therefore, this star cannot be the source of the eclipses and  has  a negligible impact on the measurement of the planet radius (a correction of around 0.002\%).
Hence, our results confirm the non-detection of close-in companions to HD~106315 within 10 arcsec presented by \citet{Crossfield2017}, whose observations have a higher sensitivity closer to the target. We also exclude the existence of significant contaminating stars present at larger distances, but inside the K2 aperture. Ignoring the presence of all possible sources inside the photometric aperture can give rise to false positives \citep{Cabrera2017}.

\section{Spectral analysis of the host star}

\label{abundances}

In order to perform the spectral analysis of the host star we first co-added all the (Doppler corrected) individual spectra of HD~106315 into a single 1D spectra with IRAF (Image Reduction and Analysis Facility).\footnote{IRAF is distributed by National Optical Astronomy Observatories, operated by the Association of Universities for Research in Astronomy, Inc., under contract with the National Science Foundation, USA.} We found that the stellar parameters derived with the equivalent width (EW) method (in our case ARES+MOOG, \citealt{Sousa2014}) were not very reliable due to the fast rotation of the host star.  In HD~106315 the fast rotation leads to a significant spectral line broadening that affects our spectral parameter retrieval method. Therefore, we adopted the values reported by \citet{Rodriguez2017} using the stellar parameter classification tool: \teff\  $= 6251 \pm 52$ K,  \logg\  = $4.1 \pm 0.1$  [cgs] and  $[$Fe/H$]$ = $- 0.27 \pm0.08$. These are in close agreement with the  \citet{Crossfield2017}  parameters. As explained in the data analysis (Section \ref{analysis}), these parameters were used as priors and fitted together with all the data we have for this system, including the modelling of the spectral energy distribution (SED) of the star and the stellar evolution tracks of Dartmouth \citep{Dotter2008}. From our combined analysis we obtained the final stellar parameters: \teff\ = 6327 $\pm$ 48\,K ,   \logg\  = 4.252  $\pm$ 4.252 0.043 [cgs]  and  $[$Fe/H$]$ =  -0.311 $\pm$ 0.079 dex. These parameters are naturally compatible with those adopted from \citet{Rodriguez2017}. 

These stellar parameters were used to derive the stellar chemical abundances, which  for  most of the elements were derived under the assumption of local  thermodynamic  equilibrium (LTE) using the 2014 version of the code MOOG \citep{Sneden1973} with the \textit{abfind} driver. For the lines affected by hyperfine splitting (HFS), we used the \textit{blends} driver. A grid of Kurucz ATLAS9 atmospheres \citep{Kurucz1993} was used as input along with the EWs and the atomic parameters, wavelength ($\lambda$), excitation energy of the lower energy level ($\chi$), and oscillator strength (log \textit{gf}) of each line. 
The EWs of the different species lines were measured automatically with the version 2 of the ARES programme\footnote{The ARES code can be downloaded at http://www.astro.up.pt/$\sim$sousasag/ares/.} \citep{Sousa2007,Sousa2015}. However, due to the high rotation of the star, some of the stellar lines were blended. The EWs of these lines were measured manually with the task \textit{splot} in IRAF.  Abundances are measured from a few isolated lines and our procedure is valid for these manually checked lines. As mentioned above, in this case the stellar parameter determination is not reliable because it is a global fit. We refer the reader to the works of \citet{Adibekyan2012}, \citet{Santos2015}, and \citet{Delgadomena2017} for further details and the complete line list. Li abundances were derived with the spectral synthesis method, also using the code MOOG and ATLAS atmosphere models as done in \citet{Delgadomena2014}.  The derived chemical abundances are given in Table~\ref{abundances}. We derive a stellar age of $3.4 \pm 2.5$ Gyr using an empirical relation with [Y/Mg] \citep{TucciMaia2016}, which is in good agreement with the age derived in Section~\ref{analysis} using stellar evolution tracks.

\begin{table}
\caption{List of the abundances derived from the HARPS combined spectra of HD~106315.\label{abundances} }
\begin{tabular}{lccc}
\hline
\hline
Element  & Abundance & Error  & Number of lines \\
 & dex & dex & \\
\hline
$[$CI/H] & -0.13  & 0.03  & 2 \\
$[$NaI/H]       & -0.13&        0.04    &2 \\
$[$MgI/H]       &-0.26& 0.11    &2 \\
$[$AlI/H]       &-0.25& 0.04    &2 \\
$[$SiI/H]       &-0.11& 0.04 &  13 \\
$[$CaI/H]       &-0.11& 0.04    &9 \\
$[$ScII/H]      &-0.29 &        0.04 &  4 \\
$[$TiI/H]       &-0.18& 0.08 &  14 \\
$[$TiII/H]      &-0.23& 0.04 &  5 \\
$[$VI   /H]   &-0.27&   0.06 &  5 \\
$[$CrI/H]       &-0.19& 0.07 &  11 \\
$[$CrII/H]      &-0.14& 0.04 &  3\\
$[$MnI/H]       &-0.19& 0.10     & 4 \\
$[$CoI/H]       &-0.24 &        0.06 &  4 \\
$[$NiI/H]       &-0.25 &        0.07 &  32 \\
$[$CuI/H]  &-0.40  &0.11  & 4 \\
$[$ZnI/H]  &-0.32  &0.1 &3 \\
$[$SrI/H]  &-0.19  &0.08  &1 \\
$[$YII/H]    & -0.13  &0.07 & 4  \\
$[$ZrII/H]   & -0.19  &0.05  &4 \\
$[$BaII/H]  & -0.08  &0.07 & 3  \\
$[$CeII/H]  & -0.22 &0.07& 3  \\
$[$NdII/H]  & -0.35  &0.11  &4\\
\hline
A(Li)*   & 2.58 &  0.08 dex & 1 \\
\hline
\hline
\end{tabular}
\\
*A(Li) = log[N(Li)/N(H)] + 12
\end{table}

\section{Data analysis}
\label{analysis}

We  jointly  analysed  the  HARPS radial velocities (RVs), the K2 light curve,  the LCO light curve  of HD~106315 and its SED as observed by the 2-MASS, Hipparcos, and wide-field infrared survey explorer (WISE)surveys \citep{Munari2014,Hog2000, Cutri2014} using the{ \sc PASTIS} software \citep{Diaz2014, Santerne2015}.  Both planets HD~106315b and HD~106315c are modelled simultaneously. We analyse only sections of the light curve centred at the mid-transit time of both planets and with a length of 2.5 transit durations.  The RVs were modelled with Keplerian orbits and the transits were modelled with the{ \sc JKTEBOP} package \citep{Southworth2008}, using an oversampling factor of 30 for K2 to account for the long integration time of the data \citep{Kipping2010}. The SED was modelled using the BT-SETTL library of stellar atmospheres  \citep{Allard2012}. This model was coupled with a Markov Chain Monte Carlo (MCMC) method  with many different parallel Metropolis-Hastings chains  \citep{Diaz2014} to derive the system parameters and their uncertainties.  At each step of the chains, the spectroscopic stellar parameters were converted into physical parameters (stellar mass and stellar radius) using the Darthmouth evolution tracks \citep{Dotter2008}. 
The coefficients of the quadratic limb-darkening law are also computed at each step of the chains using the stellar parameters and  the tables of \citet{Claret2011}.  Our choice of the limb-darkening treatment is motivated by the fact that the low cadence of the K2 data and the precision of the ground-based transit does not allow us to constrain the limb-darkening coefficients. It was shown by \citet{Muller2013} that fixing the limb-darkening coefficients to the theoretical models is adequate in most cases and can be better than fitting them, especially for low-impact parameters. Based on \citet{Espinoza2015}, we estimated that this procedure could bias the derived planet-to-star radius ratio by less than 1\%, which is below our errors.  We derive limb-darkening values of $u_1 = 0.3024 \pm 0.0049 $ and $u_2 = 0.3057 \pm 0.0018$ for the K2 light curve, and  $u_1 = 0.2270 \pm 0.004 $, $u_2=  0.3033 \pm 0.0018$ for the LCO light curve. We included a jitter parameter in the fit for each of the observation sets: K2 light curve, LCO light curve, and HARPS RVs. To model the possible effects of stellar activity in the radial velocity data, we used a Gaussian process (GP) with a quasi periodic kernel \citep{Haywood2014}. In this case the covariance matrix is defined as:

\begin{equation}
K = A^2 \exp \left[ -\frac{1}{2}  \left(\frac{\Delta t}{\lambda_1} \right)^2  -2   \left( \frac{\  \sin \left(\frac{\pi \Delta t}{P_{rot}} \right) }{\lambda_2}  \right)^2   \right]  + I \sqrt{\sigma^2 + \sigma_j^2 }.
\end{equation}
The first term is the quasi periodic kernel with the hyper-parameters:  amplitude (A),  rotation period ($P_{rot}$),  and two timescales ($\lambda_1,\lambda_2$ ). Finally,  $\Delta t $ is a matrix with elements $\Delta t_{ij} = t_i -t_j $ . The second term is the identity matrix multiplied by usual data uncertainties $\sigma$ plus a jitter term to account for uncorrelated noise  $\sigma_j$.

The priors for all the fitted parameters used in the model are given in Table~\ref{PASTISparams} in the Appendix. For the stellar distance we used a normal distribution prior centred on the {\it Gaia} parallax \citep{Gaia2016,Gaia2016b}.
 As mentioned in the previous section, for the effective temperature, the surface gravity, and metallicity we used normal priors centred on the values reported by \citet{Rodriguez2017}.  We also used normal priors for the orbital ephemeris centred on the values found by the detection pipeline and assuming a width of 0.08 d and 0.1 d for the period and transit epoch, respectively. For the orbital eccentricity, we chose a  $\beta$ distribution as prior \citep{Kipping2013} and for the planet's inclination we used a Sine distribution as prior (between 70\degree\  and 90\degree). For the remaining parameters we used uninformative priors.

For the first exploration of the parameter space, we ran the MCMC with  20 chains of $3\times 10^5$ iterations  each,  with starting  points  randomly  drawn  from  the  joint  prior. We used a Kolmogorov-Smirnov test to reject non-converging chains \citep{Diaz2014}. 
In order to better explore the region of the parameter space close to the solution and derive accurate uncertainties, we ran a second MCMC with chains starting at the best solution found from the combination of the chains that converged. For this second MCMC, we ran 20 chains of $3\times 10^5$ iterations. The burn-in phase was removed before merging the converged chains to obtain the system parameters.

\section{Results}
The parameters derived by our model are given in Table~\ref{PASTISresults}. All the uncertainties provided in the table are statistical and they do not include unknown errors in the models.  We reiterate that the stellar parameters presented in the table are derived from the combined analysis of the data and not from the spectral analysis.

We derive the radius for both planets, $ r_b =  2.44 \pm 0.17\, $ \rearth\ and $ r_c = 4.35 \pm 0.23\, $ \rearth,  which are in agreement with 
previously derived radii though closer to the values of \citet{Rodriguez2017} due to the scaling with the radius of the star. Since we used as priors the stellar parameters derived by \citet{Rodriguez2017} and a similar data analysis method, their stellar radius is very close to our derived value while the value of \citet{Crossfield2017} is slightly smaller.  Due to our high precision RVs, we were also able to derive precisely the mass of both planets:  $m_b = 12.6 \pm 3.2\, $  \mearth\  and $m_c =15.2 \pm 3.7\, $ \mearth. 
 Therefore, the inner planet is much denser (4.7 $\pm$ 1.7 g\,cm$^{-3}$ ) than the outer planet  (1.01 $\pm 0.29\,$ g\,cm$^{-3} $).
The ground-based transit obtained by  LCO allows us to update the ephemeris of HD~106315c,  decreasing the uncertainty in the period by one order of magnitude. We find P = 21.05704 $\pm$ 4.6$\times10^{-4}\,$days and T0 =  2457569.0173 $\pm$ 1.4$\times10^{-3}\,$days, which will help with future follow-up observations. We also find no significant transit timing variations.

The derived GP period is 2.825  $\pm$  0.012. This value is close to the half of the rotation period of the star derived from $v \sin i $ assuming $i=90$.   We find a similar periodicity in the FWHM indicating that this period is due to activity. However, the double of the period is not significant in the data. If this signal is due to activity, either the radial velocities show half the period of rotation of the star or the star has an inclination relative to the planetary orbit of $\sim$30 degrees. Alternatively, the variability-induced signal found in the RVs could be due to stellar super granulation, although the timescale of super granulation for this stellar type is expected to be lower.

To test the robustness of our results we redid all the analysis using the stellar parameters from \citet{Crossfield2017}. We found that all the derived values are  within $1\sigma\,$ of our final values, which are based on stellar parameters from \citet{Rodriguez2017}.
We tested the existence of a linear, quadratic, and cubic drift. Using the Bayesian information criterium, we conclude that none of these drifts are significantly supported by the current data ($\Delta BIC$ < 5 comparing with no drift ).
To further test the presence of this drift, we obtained an extra two HARPS radial velocities in early July 2017 (two months after the end of our intensive campaign) and still found no significant drift. These two points were ignored in our final analysis as they are completely uncorrelated with the intensive campaign. \citet{Crossfield2017} previously reported a hint of a drift of $0.3\pm 0.1$ m\,s$^{-1}$day$^{-1}$ ; this translates into 47 m\,s$^{-1}$ over the 157 days of the total duration of our observations. Hence, the existence of this previously reported drift is also ruled out by our observations.

\begin{table*}[h!]
\caption{Physical parameters of the HD~106315 system. \label{PASTISresults}}
\begin{center}
\begin{tabular}{l  c c}
\hline
\hline
Parameter & \multicolumn{2}{c}{Value and uncertainty}\\
\hline
{\it Orbital parameters} & {\bf HD~106315b} & {\bf HD~106315c}   \\
 &&\\
Orbital period $P$ [d]  & 9.55237 $\pm$ 8.9$\times10^{-4}$ & 21.05704 $\pm$ 4.6$\times10^{-4}$\\
Epoch of first transit T$_{0}$ [BJD$_{\rm TDB}$ - $2.4\times10^6$] & 57586.5487 $\pm$ 2.9$\times10^{-3}$ & 57569.0173 $\pm$ 1.4$\times10^{-3}$\\
Orbital eccentricity $e$ & 0.093$^{_{+0.110}}_{^{-0.068}}$  & 0.22 $\pm$ 0.15 \\
Argument of periastron $\omega$ [\degree]  & 239$^{_{+74}}_{^{-200}}$ & 96 $^{_{+88}}_{^{-35}}$ \\
Semi-major axis $a$ [AU] & 0.0907 $\pm$ 1.0 $\times10^{-3}$ & 0.1536 $\pm$ 1.7$\times10^{-3}$\\
Inclination $i$ [\degree]  & 87.54 $\pm$ 0.32 & 88.61 $^{_{+0.70}}_{^{-0.25}}$ \\
&&\\
\hline
{\it Transit \& radial velocity parameters} 
 \\
 &\\
 System scale $a/R_{\star}$ & 15.07 $\pm$ 0.70 & 25.5 $\pm$ 1.2\\
 Impact parameter $b$ & 0.67 $\pm$ 0.1 & 0.52 $^{_{+0.17}}_{^{-0.31}}$\\
 Transit duration T$_{14}$ [h] & 3.775 $\pm$ 0.081 & 4.638 $\pm$ 0.072\\
 Planet-to-star radius ratio $k_{r}$  & 0.01728 $\pm$ 6.0 $\times10^{-4}$ & 0.03086 $\pm$ 0.0010\\
 Radial velocity amplitude $K$ [\ms]  & 3.63 $\pm$ 0.92 & 3.43 $\pm$ 0.83\\
 &&\\
\hline
{\it Planet parameters} &
 \\
 &&\\
Planet mass $M_{p}$ [M$_\oplus$] & 12.6 $\pm$ 3.2 & 15.2 $\pm$ 3.7\\
Planet radius $R_{p}$ [R$_\oplus$] & 2.44 $\pm$ 0.17 & 4.35 $\pm$ 0.23\\
Planet density $\rho_{p}$ [g cm$^{-3}$] & 4.7 $\pm$ 1.7 & 1.01 $\pm$ 0.29 \\
Equilibrium temperature $T_{\rm eq}$ [K] & 1153 $\pm$ 25 & 886 $\pm$ 20 \\
&&\\
\hline
{\it Stellar parameters} \\
 &&\\
Stellar mass $M_{\star}$ [M$_\odot$] & \multicolumn{2}{c}{ 1.091 $\pm$ 0.036}\\
Stellar radius $R_{\star}$ [R$_\odot$] & \multicolumn{2}{c}{1.296 $\pm$ 0.058}\\
Stellar age $\tau$ [Gyr] &  \multicolumn{2}{c}{4.48 $\pm$ 0.96}\\
Effective temperature \teff\ [K]  & \multicolumn{2}{c}{6327 $\pm$ 48}\\
Surface gravity \logg\ [g cm$^{-2}$]  & \multicolumn{2}{c}{4.252 $\pm$ 0.043}\\
Iron abundance [Fe/H] [dex]  & \multicolumn{2}{c}{-0.311 $\pm$ 0.079}\\
Reddening E(B-V) [mag]  & \multicolumn{2}{c}{0.0050 $^{_{+0.0085}}_{^{-0.0038}}$}\\
Systemic radial velocity $\upsilon_{0}$ [\kms]  & \multicolumn{2}{c}{-3.34$^{_{+0.56}}_{^{-0.38}}$}\\  
Distance to Earth $d$ [pc] & \multicolumn{2}{c}{109 $\pm$ 5}\\
Spectral type & \multicolumn{2}{c}{F5V}\\
 &&\\
\hline
{\it Instrumental parameters}\\
 &&\\
HARPS radial velocity jitter [\ms] & \multicolumn{2}{c}{3.05 $\pm$ 0.78}\\
SED jitter [mag]  & \multicolumn{2}{c}{0.079 $\pm$ 0.020}\\
\textit{K2} jitter [ppm]  & \multicolumn{2}{c}{44 $\pm$ 2}\\
\textit{K2} contamination [ppt]  & \multicolumn{2}{c}{3.3$^{_{+3.8}}_{^{-2.3}}$}\\
\textit{K2} flux normalisation  &  \multicolumn{2}{c}{1.000001 $\pm$ 3.1 $\times 10^{-6}$}\\
LCO jitter [ppt]  & \multicolumn{2}{c}{1.424 $\pm$ 0.065}\\
LCO contamination   & \multicolumn{2}{c}{0.1 $^{_{+0.12}}_{^{-0.07}}$}\\
LCO flux normalisation  & \multicolumn{2}{c}{0.99994  $\pm$ 0.00011}\\
GP period [d] &   \multicolumn{2}{c}{ 2.825  $\pm$  0.012}\\
GP amplitude  [\ms] &   \multicolumn{2}{c}{   0.57  $\pm$  0.35} \\
GP $\lambda_1 $  [d]  & \multicolumn{2}{c}{ 652.5  $\pm$ 340 } \\
GP $\lambda_2 $   &  \multicolumn{2}{c}{ 299.0  $\pm$  260 } \\
&&\\
\hline
\hline
\end{tabular}
\\
{ We assumed R$_\odot$=695 508km, M$_\odot$=1.98842$\times10^{30}$kg, R$_\oplus$=6 378 137m, M$_\oplus$=5.9736$\times10^{24}$kg, and 1AU=149 597 870.7km.}
\end{center}
\end{table*}%

\section{Discussion and conclusions}

To probe the composition of the two transiting planets orbiting HD~106315, we acquired high precision radial velocity observations with the HARPS spectrograph. 
These observations allow us to derive a mass for both known planets orbiting HD~106315. We find that HD~106315b has a mass of  $ 12.6 \pm 3.2\, $ \mearth\ and a density of $4.7 \pm 1.7\, $ g\,cm$^{-3}$, while HD~106315c has a mass of  15.2 $\pm$ 3.7  \mearth\ and  a density of 1.01 $\pm 0.29\,$ g\,cm$^{-3}$.  This system embodies the diversity of planetary composition given that planet c is almost double the size of planet b although they have almost the same mass. Therefore, we expect they will have different compositions.

\begin{figure*}
\centering
\sidecaption 
\includegraphics[width=12cm]{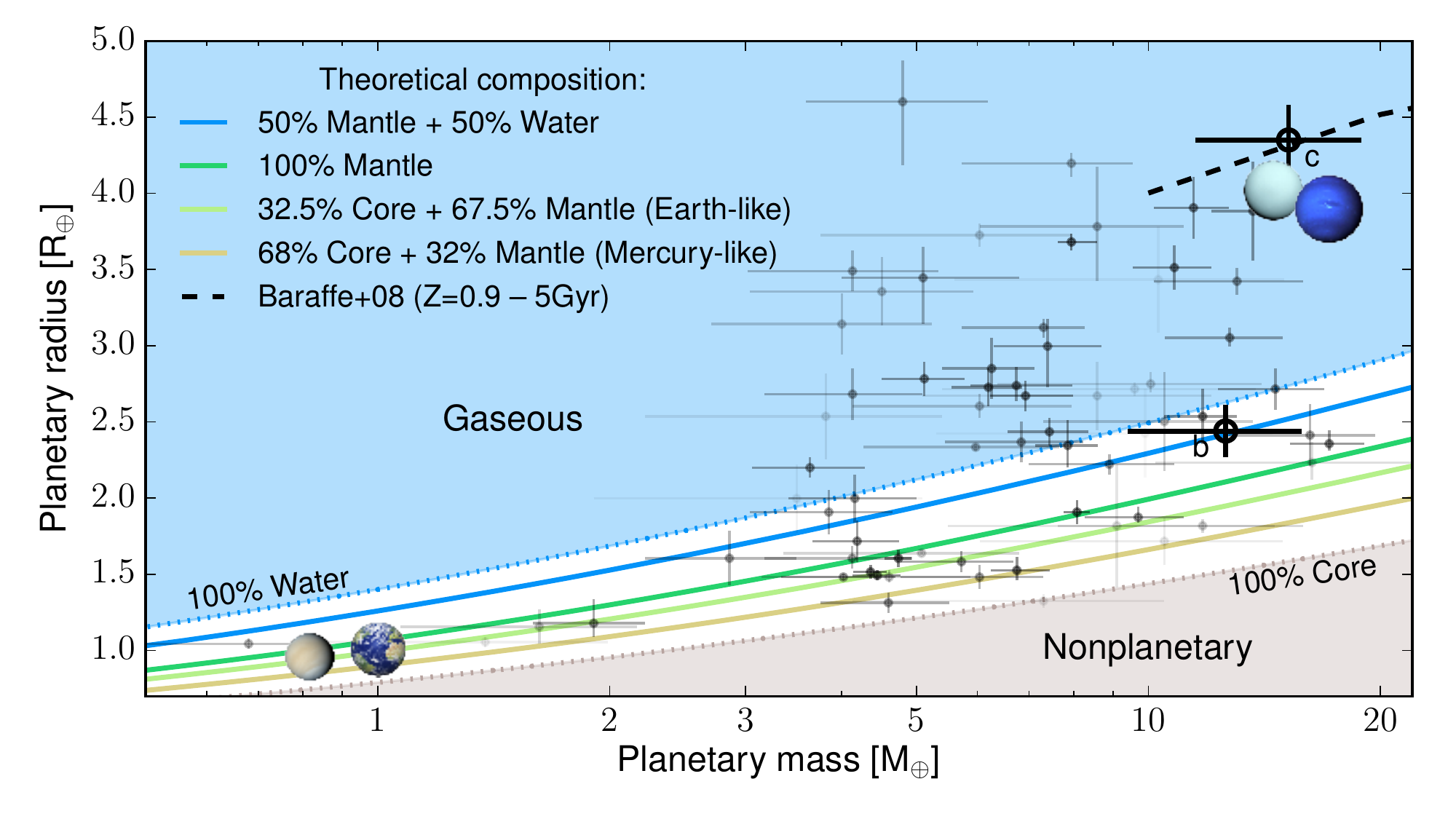}
\caption{Mass-radius relationship for different bulk compositions of small planets. We overplot different compositions for solid planets in between pure iron and pure water. The black dashed line corresponds to the models of \citet{Baraffe2008} for gaseous planets with heavy material enrichments of 0.9 and an age of 5~Gyr. We superimposed the known planets in this mass-radius range where the greyscale depends on the precision of the mass and radius. We mark the position of the planets HD~106315b and HD~106315c, as derived by us.  }
 \label{fig.massradius}  
\end{figure*}

In Figure~\ref{fig.massradius}, we show the positions of HD 106315b and HD 106315c on the mass-radius diagram compared with known planets within the same mass and size range (M~$<$~20~$\mearth$ and R~$<$~5~$\rearth$). To probe the planets' composition, we plot in the same figure the theoretical models for solid planets with assumed compositions of pure iron, pure silicate, and pure water, as well as Earth-like and Mercury-like compositions (Brugger et al. submitted). We also plot the model of \citet{Baraffe2008}, which applies to planets with gaseous envelopes with a heavy material enrichment of 0.9 and an age of 5~Gyr, which appears to be a good match for HD 106315c. HD 106315b appears to be composed of a large fraction of silicate rocks and water.

\begin{figure*}[h]
        \begin{center}
                \includegraphics[width=\textwidth]{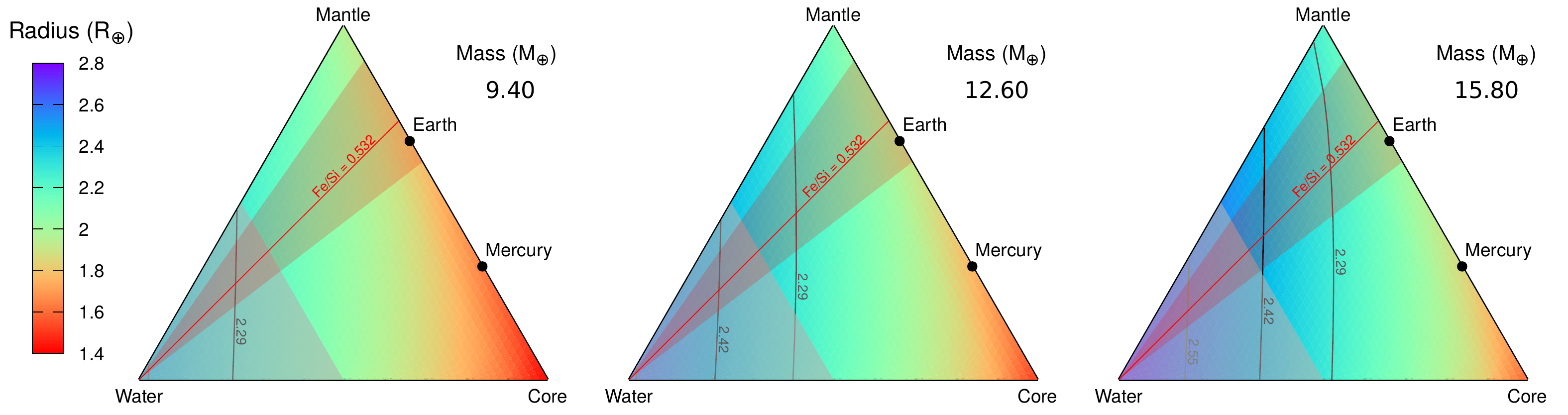}
                \caption{Ternary diagrams displaying the investigated compositional parameter space of HD~106315b for the minimum, central, and maximum masses, using 1$\sigma$ uncertainties. Also shown are the isoradius curves denoting the planet radius with the 1$\sigma$ extreme values. The planetary Fe/Si ratio assumed for HD 106315b, with its associated 1$\sigma$ uncertainties, delimits an area represented as a red triangle. Planetary compositions with a water mass fraction higher than 50\% are excluded, based on assumptions on the solar system's present properties (grey zone).}
                \label{fig:ternary}
        \end{center}
\end{figure*}

A detailed investigation of HD 106315b's composition using the planetary interior model by Brugger (et al. submitted) allows us to constrain the core mass and potential water content of this planet. 
This model includes only refractory elements, and hence it does not consider the possibility of a rocky core with a thick H-He envelope. Therefore, we cannot apply this model to planet c because, as discussed above, planet c must have an envelop of H-He.  The model assumes that the planet is fully differentiated into three main layers: a metallic core, a silicate mantle, and a water envelope. The size, and thus mass, of these layers fix a planet's internal structure, that is, its composition. In this model, the surface conditions of HD 106315b are assumed to be close to those of the Earth (288~K temperature and 1~bar pressure) to allow the simulation of liquid water. Considering the equilibrium temperature inferred for HD 106315b (1153~$\pm$~25~K), water on the surface of this planet would be in the supercritical phase. We explore all possible compositions for HD 106315b, with a parameter space that is represented by ternary diagrams for the core, mantle, and water mass fractions (Figure~\ref{fig:ternary}). Our simulations show that the mass and radius measured for HD 106315b are incompatible with a fully rocky composition, as the water mass fraction of this planet has a lower limit of 5\%. This fraction may theoretically go up to 100\% but we can place an upper limit of 50\%, which is the maximum value found in large differentiated bodies of the solar system. The core mass fraction of HD 106315b is found to be in the range 0--52\%. Given the uncertainties on the fundamental parameters of this planet, its core mass and water fraction cannot be better constrained. However, the incorporation of the planet's bulk Fe/Si ratio helps to reduce the computed ranges under the assumption that this ratio reflects that of the host star \citep{Thiabaud2015}. Using the stellar abundances from Section~3, we compute that Fe/Si = 0.532~$\pm$~0.316. Here, incorporating this parameter in the interior model gives  a 9-50\% range for the water mass fraction and  5-29\% for the core mass fraction. It appears that HD 106315b cannot be fully rocky and must harbour a significant water envelope. Both the water mass fraction and the core mass fraction of this planet can be significantly constrained via the use of the stellar Fe/Si ratio and the limitations from solar system formation conditions, respectively.

The other possibility is that HD~106315b is composed of a rocky core with a thick H-He envelope. This could be expected since HD 106315b is close to the transition between rocky and gaseous planets. HD 106315b and c have the potential to bring light to this transition, given their close masses but different radii.

To probe the evolution of this interesting multi-planetary system, we compared our derived systems parameters with the results of synthetic planetary populations and evolutionary tracks calculated with the Bern model of planet formation and evolution \citep{Alibert2005, Mordasini2012,Benz2014} available through DACE \footnote{https://dace.unige.ch.} (Data and Analysis Center for Exoplanets). The simulations assume a stellar mass of 1 \Msun\ , a stellar age of $4$Gyr, and cold gas accretion.  The long-term evolution neglects envelope evaporation. We compared simulations with similar semi-major axis and mass to the values we derived for both the planets. All simulations have a semi-major axis higher than 0.101 AU. Therefore they cannot reproduce planet b. However, using the simulation with semi-major axis  $\sim$0.101 AU as proxy for planet b, we find that this planet could have been formed in between 0.11 and 2.5 AU and probably has a core mass just slightly lower ($\sim$12\mearth) than the total planetary mass. We found a close proxy for planet c, which suggests that it could have been formed between 1.0 and 3.0 AU. Hence, in general planet c formed further away from the host star than planet b and is expected to have similar core mass ($\sim$ 12 \mearth). Another interesting fact is that the predicted radius for both planets is almost the same and in agreement with the value we measured for planet c. However,  the radii of the simulated planets are always higher than the radius of planet b. This could be explained if planet b, as it lies closer to the host star,  has suffered envelope evaporation, which is not taken into account in the simulation. Therefore, the difference in the density between both planets can be explained by the different formation region in the proto-planetary disc as well as a possible envelope evaporation. 

With Equation 15 of \citet{Lecavelier2007}, we estimate the escape rate of the atmosphere using two different values for the $F_{EUV}$ (1AU):  14.7 erg cm$^{-2}$s$^{-1}$ from  \citet{Lecavelier2007} and 82.76  erg cm$^{-2}$s$^{-1}$ calculated using Equation 3 of \citet{Ehrenreich2011}. Assuming that the $F_{EUV}$ (1AU)  is constant, we calculate an atmosphere loss of  $\sim $ [  0.018, 0.10] earth masses for planet b and $\sim$  [0.033, 0.19] for planet c during their lifetime of 4 Gyr. Therefore, it appears that the mass loss is not important for these two planets. However, our first approximation of the mass loss rates should be confirmed with more detailed models of evaporating atmospheres, which are out of the scope of this paper.

Gaining insight into the composition of the atmosphere of these planets would allow us to better constrain their interior composition and their formation history.
We estimate the scale height of the atmosphere of HD~106315b and HD~106315c to be $\sim$196.72 Km and  $\sim$ 464.53 Km, respectively. Therefore,  given the brightness of the host star, both planets are also golden targets for transmission spectroscopy with JWST. This would be particularly interesting for HD~106315b since this planet lies in the transition between rocky and gaseous planet composition. Probing its atmosphere with upcoming new facilities like JWST and the extremely large telescopes (ELTs) will help to better understand the composition of this planet.

\begin{acknowledgements}
We are grateful to the pool of HARPS observers who conducted part of the visitor-mode observations at La Silla Observatory: Romina Ibanez Bustos, Nicola Astudillo, Aur\'elien Wyttenbach, Esther Linder, Xavier Bonfils, Elodie H\'ebrard, and Alejandro Suarez. This work makes use of observations from the LCO network.
This publication makes use of The Data \& Analysis Center for Exoplanets (DACE), which is a facility based at the University of Geneva (CH) dedicated to extrasolar planets' data visualisation, exchange, and analysis. DACE is a platform of the Swiss National Centre of Competence in Research (NCCR) PlanetS, federating the Swiss expertise in Exoplanet research. The DACE platform is available at https://dace.unige.ch. This research has made use of the NASA Exoplanet Archive, which is operated by the California Institute of Technology, under contract with the National Aeronautics and Space Administration under the Exoplanet Exploration Program. This research has made use of the VizieR catalogue access tool, CDS, Strasbourg, France. The original description of the VizieR service was published in A\&AS 143, 23. This publication makes use of data products from the Two Micron All Sky Survey, which is a joint project of the University of Massachusetts and the Infrared Processing and Analysis Center/California Institute of Technology, funded by the National Aeronautics and Space Administration and the National Science Foundation. This publication makes use of data products from the Wide-field Infrared Survey Explorer, which is a joint project of the University of California, Los Angeles, and the Jet Propulsion Laboratory/California Institute of Technology, funded by the National Aeronautics and Space Administration.
The IA/Porto team acknowledges support from Funda\c{c}\~ao para a Ci\^encia e a Tecnologia (FCT) through national funds
and by FEDER through COMPETE2020 by these grants UID/FIS/04434/2013 \& POCI-01-0145-FEDER-007672  , PTDC/FIS-AST/1526/2014 \&  POCI-01-0145-FEDER-016886 and  PTDC/FIS-AST/7073/2014 \& POCI-01-0145-FEDER-016880. SCCB, E.D.M. V.Zh.A., N.C.S., PF, and S.G.S. also acknowledge  support from FCT through Investigador FCT contracts IF/01312/2014/CP1215/CT000,  IF/00849/2015,  IF/00650/2015/CP1273/CT0001, IF/00169/2012/CP0150/CT0002,  IF/01037/2013/CP1191/CT0001 and IF/00028/2014/CP1215/CT0002 funded by FCT (Portugal) and POPH/FSE (EC). 
VZhA and JPF also acknowledge support from the FCT in the form of the grants SFRH/BPD/70574/2010 and SFRH/BD/93848/2013, respectively. PF further acknowledges  support from Funda\c{c}\~ao para a Ci\^encia e a Tecnologia (FCT)  through POPH/FSE (EC) by FEDER funding through the programme ``Programa Operacional de Factores de Competitividade - COMPETE'' and exploratory project of reference IF/01037/2013/CP1191/CT0001. J.P.F., S.H., and J.J.N. acknowledge support by the fellowships SFRH/BD/93848/2013, PD/BD/128119/2016, and PD/BD/52700/2014, funded by FCT (Portugal) and POPH/FSE (EC). The French group acknowledges financial support from the French Programme National de Plan\'etologie (PNP, INSU). The Swiss group acknowledges financial support the National Centre for Competence in Research "PlanetS" supported by the Swiss National Science Foundation (SNSF).
DJA is funded under STFC consolidated grant reference ST/P000495/1. This work has been partly carried out thanks to support funding from Excellence Initiative of Aix-Marseille University A*MIDEX, a French "Investissements d'Avenir" programme.  ACC acknowledges support from STFC consolidated  grant number ST/M001296/1. XD is grateful to the Society in Science? The Branco Weiss Fellowship for its financial support. Part of this research has been funded by the Spanish grant  ESP2015-65712-C5-1-R.
We thank the anonymous referee for the careful review of this manuscript that improved its quality.
\end{acknowledgements}

\bibliographystyle{aa} 
\bibliography{susana}

\begin{thebibliography}{86}
\expandafter\ifx\csname natexlab\endcsname\relax\def\natexlab#1{#1}\fi

\bibitem[{{Adibekyan} {et~al.}(2012){Adibekyan}, {Sousa}, {Santos}, {Delgado
  Mena}, {Gonz{\'a}lez Hern{\'a}ndez}, {Israelian}, {Mayor}, \&
  {Khachatryan}}]{Adibekyan2012}
{Adibekyan}, V.~Z., {Sousa}, S.~G., {Santos}, N.~C., {et~al.} 2012, \aap, 545,
  A32

\bibitem[{{Albrecht} {et~al.}(2012){Albrecht}, {Winn}, {Johnson}, {Howard},
  {Marcy}, {Butler}, {Arriagada}, {Crane}, {Shectman}, {Thompson}, {Hirano},
  {Bakos}, \& {Hartman}}]{Albrecht2012}
{Albrecht}, S., {Winn}, J.~N., {Johnson}, J.~A., {et~al.} 2012, \apj, 757, 18

\bibitem[{{Alibert}(2014)}]{Alibert2014}
{Alibert}, Y. 2014, \aap, 561, A41

\bibitem[{{Alibert} {et~al.}(2005){Alibert}, {Mordasini}, {Benz}, \&
  {Winisdoerffer}}]{Alibert2005}
{Alibert}, Y., {Mordasini}, C., {Benz}, W., \& {Winisdoerffer}, C. 2005, \aap,
  434, 343

\bibitem[{{Allard} {et~al.}(2012){Allard}, {Homeier}, \&
  {Freytag}}]{Allard2012}
{Allard}, F., {Homeier}, D., \& {Freytag}, B. 2012, Royal Society of London
  Philosophical Transactions Series A, 370, 2765

\bibitem[{{Baraffe} {et~al.}(2008){Baraffe}, {Chabrier}, \&
  {Barman}}]{Baraffe2008}
{Baraffe}, I., {Chabrier}, G., \& {Barman}, T. 2008, \aap, 482, 315

\bibitem[{{Baranne} {et~al.}(1996){Baranne}, {Queloz}, {Mayor}, {Adrianzyk},
  {Knispel}, {Kohler}, {Lacroix}, {Meunier}, {Rimbaud}, \& {Vin}}]{Baranne1996}
{Baranne}, A., {Queloz}, D., {Mayor}, M., {et~al.} 1996, \aaps, 119, 373

\bibitem[{{Barclay} {et~al.}(2013){Barclay}, {Rowe}, {Lissauer}, {Huber},
  {Fressin}, {Howell}, {Bryson}, {Chaplin}, {D{\'e}sert}, {Lopez}, {Marcy},
  {Mullally}, {Ragozzine}, {Torres}, {Adams}, {Agol}, {Barrado}, {Basu},
  {Bedding}, {Buchhave}, {Charbonneau}, {Christiansen},
  {Christensen-Dalsgaard}, {Ciardi}, {Cochran}, {Dupree}, {Elsworth},
  {Everett}, {Fischer}, {Ford}, {Fortney}, {Geary}, {Haas}, {Handberg},
  {Hekker}, {Henze}, {Horch}, {Howard}, {Hunter}, {Isaacson}, {Jenkins},
  {Karoff}, {Kawaler}, {Kjeldsen}, {Klaus}, {Latham}, {Li}, {Lillo-Box},
  {Lund}, {Lundkvist}, {Metcalfe}, {Miglio}, {Morris}, {Quintana}, {Stello},
  {Smith}, {Still}, \& {Thompson}}]{Barclay2013}
{Barclay}, T., {Rowe}, J.~F., {Lissauer}, J.~J., {et~al.} 2013, \nat, 494, 452

\bibitem[{{Barros} {et~al.}(2014){Barros}, {Almenara}, {Deleuil}, {Diaz},
  {Csizmadia}, {Cabrera}, {Chaintreuil}, {Collier Cameron}, {Hatzes},
  {Haywood}, {Lanza}, {Aigrain}, {Alonso}, {Bord{\'e}}, {Bouchy}, {Deeg},
  {Erikson}, {Fridlund}, {Grziwa}, {Gandolfi}, {Guillot}, {Guenther}, {Leger},
  {Moutou}, {Ollivier}, {Pasternacki}, {P{\"a}tzold}, {Rauer}, {Rouan},
  {Santerne}, {Schneider}, \& {Wuchterl}}]{Barros2014}
{Barros}, S.~C.~C., {Almenara}, J.~M., {Deleuil}, M., {et~al.} 2014, \aap, 569,
  A74

\bibitem[{{Barros} {et~al.}(2016){Barros}, {Demangeon}, \&
  {Deleuil}}]{Barros2016}
{Barros}, S.~C.~C., {Demangeon}, O., \& {Deleuil}, M. 2016, \aap, 594, A100

\bibitem[{{Batalha} {et~al.}(2017){Batalha}, {Kempton}, \&
  {Mbarek}}]{Batalha2017}
{Batalha}, N.~E., {Kempton}, E.~M.-R., \& {Mbarek}, R. 2017, \apjl, 836, L5

\bibitem[{{Benz} {et~al.}(2014){Benz}, {Ida}, {Alibert}, {Lin}, \&
  {Mordasini}}]{Benz2014}
{Benz}, W., {Ida}, S., {Alibert}, Y., {Lin}, D., \& {Mordasini}, C. 2014,
  Protostars and Planets VI, 691

\bibitem[{{Bonomo} {et~al.}(2010){Bonomo}, {Santerne}, {Alonso}, {Gazzano},
  {Havel}, {Aigrain}, {Auvergne}, {Baglin}, {Barbieri}, {Barge}, {Benz},
  {Bord{\'e}}, {Bouchy}, {Bruntt}, {Cabrera}, {Collier Cameron}, {Carone},
  {Carpano}, {Csizmadia}, {Deleuil}, {Deeg}, {Dvorak}, {Erikson},
  {Ferraz-Mello}, {Fridlund}, {Gandolfi}, {Gillon}, {Guenther}, {Guillot},
  {Hatzes}, {H{\'e}brard}, {Jorda}, {Lammer}, {Lanza}, {L{\'e}ger}, {Llebaria},
  {Mayor}, {Mazeh}, {Moutou}, {Ollivier}, {P{\"a}tzold}, {Pepe}, {Queloz},
  {Rauer}, {Rouan}, {Samuel}, {Schneider}, {Tingley}, {Udry}, \&
  {Wuchterl}}]{Bonomo2010}
{Bonomo}, A.~S., {Santerne}, A., {Alonso}, R., {et~al.} 2010, \aap, 520, A65

\bibitem[{{Borucki} {et~al.}(2010){Borucki}, {Koch}, {Basri}, {Batalha},
  {Brown}, {Caldwell}, {Caldwell}, {Christensen-Dalsgaard}, {Cochran},
  {DeVore}, {Dunham}, {Dupree}, {Gautier}, {Geary}, {Gilliland}, {Gould},
  {Howell}, {Jenkins}, {Kondo}, {Latham}, {Marcy}, {Meibom}, {Kjeldsen},
  {Lissauer}, {Monet}, {Morrison}, {Sasselov}, {Tarter}, {Boss}, {Brownlee},
  {Owen}, {Buzasi}, {Charbonneau}, {Doyle}, {Fortney}, {Ford}, {Holman},
  {Seager}, {Steffen}, {Welsh}, {Rowe}, {Anderson}, {Buchhave}, {Ciardi},
  {Walkowicz}, {Sherry}, {Horch}, {Isaacson}, {Everett}, {Fischer}, {Torres},
  {Johnson}, {Endl}, {MacQueen}, {Bryson}, {Dotson}, {Haas}, {Kolodziejczak},
  {Van Cleve}, {Chandrasekaran}, {Twicken}, {Quintana}, {Clarke}, {Allen},
  {Li}, {Wu}, {Tenenbaum}, {Verner}, {Bruhweiler}, {Barnes}, \&
  {Prsa}}]{Borucki2010}
{Borucki}, W.~J., {Koch}, D., {Basri}, G., {et~al.} 2010, Science, 327, 977

\bibitem[{{Bouchy} {et~al.}(2004){Bouchy}, {Maeder}, {Mayor}, {M{\'e}gevand},
  {Pepe}, \& {Sosnowska}}]{Bouchy2004}
{Bouchy}, F., {Maeder}, A., {Mayor}, M., {et~al.} 2004, \nat, 432, 2

\bibitem[{{Bouchy} {et~al.}(2001){Bouchy}, {Pepe}, \& {Queloz}}]{Bouchy2001}
{Bouchy}, F., {Pepe}, F., \& {Queloz}, D. 2001, \aap, 374, 733

\bibitem[{{Brown} {et~al.}(2013){Brown}, {Baliber}, {Bianco}, {Bowman},
  {Burleson}, {Conway}, {Crellin}, {Depagne}, {De Vera}, {Dilday}, {Dragomir},
  {Dubberley}, {Eastman}, {Elphick}, {Falarski}, {Foale}, {Ford}, {Fulton},
  {Garza}, {Gomez}, {Graham}, {Greene}, {Haldeman}, {Hawkins}, {Haworth},
  {Haynes}, {Hidas}, {Hjelstrom}, {Howell}, {Hygelund}, {Lister}, {Lobdill},
  {Martinez}, {Mullins}, {Norbury}, {Parrent}, {Paulson}, {Petry}, {Pickles},
  {Posner}, {Rosing}, {Ross}, {Sand}, {Saunders}, {Shobbrook}, {Shporer},
  {Street}, {Thomas}, {Tsapras}, {Tufts}, {Valenti}, {Vander Horst}, {Walker},
  {White}, \& {Willis}}]{Brown2013}
{Brown}, T.~M., {Baliber}, N., {Bianco}, F.~B., {et~al.} 2013, \pasp, 125, 1031

\bibitem[{{Cabrera} {et~al.}(2017){Cabrera}, {Barros}, {Armstrong}, {Hidalgo},
  {Santos}, {Almenara}, {Alonso}, {Deleuil}, {Demangeon}, {Diaz}, {Lendl},
  {Pfaff}, {Rauer}, {Santerne}, {Serrano}, \& {Zucker}}]{Cabrera2017}
{Cabrera}, J., {Barros}, S.~C.~C., {Armstrong}, D., {et~al.} 2017, ArXiv
  e-prints [\eprint[arXiv]{1707.08007}]

\bibitem[{{Carter} {et~al.}(2012){Carter}, {Agol}, {Chaplin}, {Basu},
  {Bedding}, {Buchhave}, {Christensen-Dalsgaard}, {Deck}, {Elsworth},
  {Fabrycky}, {Ford}, {Fortney}, {Hale}, {Handberg}, {Hekker}, {Holman},
  {Huber}, {Karoff}, {Kawaler}, {Kjeldsen}, {Lissauer}, {Lopez}, {Lund},
  {Lundkvist}, {Metcalfe}, {Miglio}, {Rogers}, {Stello}, {Borucki}, {Bryson},
  {Christiansen}, {Cochran}, {Geary}, {Gilliland}, {Haas}, {Hall}, {Howard},
  {Jenkins}, {Klaus}, {Koch}, {Latham}, {MacQueen}, {Sasselov}, {Steffen},
  {Twicken}, \& {Winn}}]{Carter2012}
{Carter}, J.~A., {Agol}, E., {Chaplin}, W.~J., {et~al.} 2012, Science, 337, 556

\bibitem[{{Claret} \& {Bloemen}(2011)}]{Claret2011}
{Claret}, A. \& {Bloemen}, S. 2011, \aap, 529, A75+

\bibitem[{{Crossfield} {et~al.}(2017){Crossfield}, {Ciardi}, {Isaacson},
  {Howard}, {Petigura}, {Weiss}, {Fulton}, {Sinukoff}, {Schlieder}, {Mawet},
  {Ruane}, {de Pater}, {de Kleer}, {Davies}, {Christiansen}, {Dressing},
  {Hirsch}, {Benneke}, {Crepp}, {Kosiarek}, {Livingston}, {Gonzales},
  {Beichman}, \& {Knutson}}]{Crossfield2017}
{Crossfield}, I.~J.~M., {Ciardi}, D.~R., {Isaacson}, H., {et~al.} 2017, \aj,
  153, 255

\bibitem[{{Cutri}(2014)}]{Cutri2014}
{Cutri}, R.~M. 2014, VizieR Online Data Catalog, 2328

\bibitem[{{Delgado Mena} {et~al.}(2014){Delgado Mena}, {Israelian},
  {Gonz{\'a}lez Hern{\'a}ndez}, {Sousa}, {Mortier}, {Santos}, {Adibekyan},
  {Fernandes}, {Rebolo}, {Udry}, \& {Mayor}}]{Delgadomena2014}
{Delgado Mena}, E., {Israelian}, G., {Gonz{\'a}lez Hern{\'a}ndez}, J.~I.,
  {et~al.} 2014, \aap, 562, A92

\bibitem[{{Delgado Mena} {et~al.}(2017){Delgado Mena}, {Tsantaki}, {Adibekyan},
  {Sousa}, {Santos}, {Gonz{\'a}lez Hern{\'a}ndez}, \&
  {Israelian}}]{Delgadomena2017}
{Delgado Mena}, E., {Tsantaki}, M., {Adibekyan}, V.~Z., {et~al.} 2017, ArXiv
  e-prints [\eprint[arXiv]{1705.04349}]

\bibitem[{{D{\'{\i}}az} {et~al.}(2014){D{\'{\i}}az}, {Almenara}, {Santerne},
  {Moutou}, {Lethuillier}, \& {Deleuil}}]{Diaz2014}
{D{\'{\i}}az}, R.~F., {Almenara}, J.~M., {Santerne}, A., {et~al.} 2014, \mnras,
  441, 983

\bibitem[{{Dotter} {et~al.}(2008){Dotter}, {Chaboyer}, {Jevremovi{\'c}},
  {Kostov}, {Baron}, \& {Ferguson}}]{Dotter2008}
{Dotter}, A., {Chaboyer}, B., {Jevremovi{\'c}}, D., {et~al.} 2008, \apjs, 178,
  89

\bibitem[{{Dressing} {et~al.}(2015){Dressing}, {Charbonneau}, {Dumusque},
  {Gettel}, {Pepe}, {Collier Cameron}, {Latham}, {Molinari}, {Udry}, {Affer},
  {Bonomo}, {Buchhave}, {Cosentino}, {Figueira}, {Fiorenzano}, {Harutyunyan},
  {Haywood}, {Johnson}, {Lopez-Morales}, {Lovis}, {Malavolta}, {Mayor},
  {Micela}, {Motalebi}, {Nascimbeni}, {Phillips}, {Piotto}, {Pollacco},
  {Queloz}, {Rice}, {Sasselov}, {S{\'e}gransan}, {Sozzetti}, {Szentgyorgyi}, \&
  {Watson}}]{Dressing2015b}
{Dressing}, C.~D., {Charbonneau}, D., {Dumusque}, X., {et~al.} 2015, \apj, 800,
  135

\bibitem[{{Dumusque} {et~al.}(2014){Dumusque}, {Bonomo}, {Haywood},
  {Malavolta}, {S{\'e}gransan}, {Buchhave}, {Collier Cameron}, {Latham},
  {Molinari}, {Pepe}, {Udry}, {Charbonneau}, {Cosentino}, {Dressing},
  {Figueira}, {Fiorenzano}, {Gettel}, {Harutyunyan}, {Horne}, {Lopez-Morales},
  {Lovis}, {Mayor}, {Micela}, {Motalebi}, {Nascimbeni}, {Phillips}, {Piotto},
  {Pollacco}, {Queloz}, {Rice}, {Sasselov}, {Sozzetti}, {Szentgyorgyi}, \&
  {Watson}}]{Dumusque2014}
{Dumusque}, X., {Bonomo}, A.~S., {Haywood}, R.~D., {et~al.} 2014, \apj, 789,
  154

\bibitem[{{Dumusque} {et~al.}(2011){Dumusque}, {Udry}, {Lovis}, {Santos}, \&
  {Monteiro}}]{Dumusque2011}
{Dumusque}, X., {Udry}, S., {Lovis}, C., {Santos}, N.~C., \& {Monteiro},
  M.~J.~P.~F.~G. 2011, \aap, 525, A140

\bibitem[{{Ehrenreich} \& {D{\'e}sert}(2011)}]{Ehrenreich2011}
{Ehrenreich}, D. \& {D{\'e}sert}, J.-M. 2011, \aap, 529, A136

\bibitem[{{Espinoza} \& {Jord{\'a}n}(2015)}]{Espinoza2015}
{Espinoza}, N. \& {Jord{\'a}n}, A. 2015, \mnras, 450, 1879

\bibitem[{{Ford} \& {Rasio}(2008)}]{Ford2008}
{Ford}, E.~B. \& {Rasio}, F.~A. 2008, \apj, 686, 621

\bibitem[{{Fressin} {et~al.}(2013){Fressin}, {Torres}, {Charbonneau}, {Bryson},
  {Christiansen}, {Dressing}, {Jenkins}, {Walkowicz}, \&
  {Batalha}}]{Fressin2013}
{Fressin}, F., {Torres}, G., {Charbonneau}, D., {et~al.} 2013, \apj, 766, 81

\bibitem[{{Fried}(1978)}]{fried78}
{Fried}, D.~L. 1978, Journal of the Optical Society of America (1917-1983), 68,
  1651

\bibitem[{{Fulton} {et~al.}(2017){Fulton}, {Petigura}, {Howard}, {Isaacson},
  {Marcy}, {Cargile}, {Hebb}, {Weiss}, {Johnson}, {Morton}, {Sinukoff},
  {Crossfield}, \& {Hirsch}}]{Fulton2017}
{Fulton}, B.~J., {Petigura}, E.~A., {Howard}, A.~W., {et~al.} 2017, ArXiv
  e-prints [\eprint[arXiv]{1703.10375}]

\bibitem[{{Gaia Collaboration} {et~al.}(2016{\natexlab{a}}){Gaia
  Collaboration}, {Brown}, {Vallenari}, {Prusti}, {de Bruijne}, {Mignard},
  {Drimmel}, {Babusiaux}, {Bailer-Jones}, {Bastian}, \& et~al.}]{Gaia2016b}
{Gaia Collaboration}, {Brown}, A.~G.~A., {Vallenari}, A., {et~al.}
  2016{\natexlab{a}}, \aap, 595, A2

\bibitem[{{Gaia Collaboration} {et~al.}(2016{\natexlab{b}}){Gaia
  Collaboration}, {Prusti}, {de Bruijne}, {Brown}, {Vallenari}, {Babusiaux},
  {Bailer-Jones}, {Bastian}, {Biermann}, {Evans}, \& et~al.}]{Gaia2016}
{Gaia Collaboration}, {Prusti}, T., {de Bruijne}, J.~H.~J., {et~al.}
  2016{\natexlab{b}}, \aap, 595, A1

\bibitem[{{Green} {et~al.}(2015){Green}, {Schlafly}, {Finkbeiner}, {Rix},
  {Martin}, {Burgett}, {Draper}, {Flewelling}, {Hodapp}, {Kaiser}, {Kudritzki},
  {Magnier}, {Metcalfe}, {Price}, {Tonry}, \& {Wainscoat}}]{Green2015}
{Green}, G.~M., {Schlafly}, E.~F., {Finkbeiner}, D.~P., {et~al.} 2015, \apj,
  810, 25

\bibitem[{{Haywood} {et~al.}(2014){Haywood}, {Collier Cameron}, {Queloz},
  {Barros}, {Deleuil}, {Fares}, {Gillon}, {Lanza}, {Lovis}, {Moutou}, {Pepe},
  {Pollacco}, {Santerne}, {S{\'e}gransan}, \& {Unruh}}]{Haywood2014}
{Haywood}, R.~D., {Collier Cameron}, A., {Queloz}, D., {et~al.} 2014, \mnras,
  443, 2517

\bibitem[{{H{\o}g} {et~al.}(2000){H{\o}g}, {Fabricius}, {Makarov}, {Urban},
  {Corbin}, {Wycoff}, {Bastian}, {Schwekendiek}, \& {Wicenec}}]{Hog2000}
{H{\o}g}, E., {Fabricius}, C., {Makarov}, V.~V., {et~al.} 2000, \aap, 355, L27

\bibitem[{{Hormuth} {et~al.}(2008){Hormuth}, {Brandner}, {Hippler}, \&
  {Henning}}]{hormuth08}
{Hormuth}, F., {Brandner}, W., {Hippler}, S., \& {Henning}, T. 2008, Journal of
  Physics Conference Series, 131, 012051

\bibitem[{{Howell} {et~al.}(2014){Howell}, {Sobeck}, {Haas}, {Still},
  {Barclay}, {Mullally}, {Troeltzsch}, {Aigrain}, {Bryson}, {Caldwell},
  {Chaplin}, {Cochran}, {Huber}, {Marcy}, {Miglio}, {Najita}, {Smith},
  {Twicken}, \& {Fortney}}]{Howell2014}
{Howell}, S.~B., {Sobeck}, C., {Haas}, M., {et~al.} 2014, \pasp, 126, 398

\bibitem[{{Jontof-Hutter} {et~al.}(2015){Jontof-Hutter}, {Rowe}, {Lissauer},
  {Fabrycky}, \& {Ford}}]{Jontof-Hutter2015}
{Jontof-Hutter}, D., {Rowe}, J.~F., {Lissauer}, J.~J., {Fabrycky}, D.~C., \&
  {Ford}, E.~B. 2015, \nat, 522, 321

\bibitem[{{Kipping}(2010)}]{Kipping2010}
{Kipping}, D.~M. 2010, \mnras, 408, 1758

\bibitem[{{Kipping}(2013)}]{Kipping2013}
{Kipping}, D.~M. 2013, \mnras, 434, L51

\bibitem[{{Kurucz}(1993)}]{Kurucz1993}
{Kurucz}, R. 1993, ATLAS9 Stellar Atmosphere Programs and 2 km/s grid.~Kurucz
  CD-ROM No.~13.~ Cambridge, Mass.: Smithsonian Astrophysical Observatory,
  1993., 13

\bibitem[{{Lecavelier Des Etangs}(2007)}]{Lecavelier2007}
{Lecavelier Des Etangs}, A. 2007, \aap, 461, 1185

\bibitem[{{L{\'e}ger} {et~al.}(2004){L{\'e}ger}, {Selsis}, {Sotin}, {Guillot},
  {Despois}, {Mawet}, {Ollivier}, {Lab{\`e}que}, {Valette}, {Brachet},
  {Chazelas}, \& {Lammer}}]{Leger2004}
{L{\'e}ger}, A., {Selsis}, F., {Sotin}, C., {et~al.} 2004, \icarus, 169, 499

\bibitem[{{Lillo-Box} {et~al.}(2014){Lillo-Box}, {Barrado}, \&
  {Bouy}}]{lillo-box14b}
{Lillo-Box}, J., {Barrado}, D., \& {Bouy}, H. 2014, \aap, 566, A103

\bibitem[{{Lissauer}(1993)}]{Lissauer1993}
{Lissauer}, J.~J. 1993, \araa, 31, 129

\bibitem[{{Lissauer} {et~al.}(2011){Lissauer}, {Fabrycky}, {Ford}, {Borucki},
  {Fressin}, {Marcy}, {Orosz}, {Rowe}, {Torres}, {Welsh}, {Batalha}, \& {et
  al.}}]{Lissauer2011a}
{Lissauer}, J.~J., {Fabrycky}, D.~C., {Ford}, E.~B., {et~al.} 2011, \nat, 470,
  53

\bibitem[{{Lovis} {et~al.}(2011){Lovis}, {Dumusque}, {Santos}, {Bouchy},
  {Mayor}, {Pepe}, {Queloz}, {S{\'e}gransan}, \& {Udry}}]{Lovis2011}
{Lovis}, C., {Dumusque}, X., {Santos}, N.~C., {et~al.} 2011, ArXiv e-prints
  [\eprint[arXiv]{1107.5325}]

\bibitem[{{Malavolta} {et~al.}(2017){Malavolta}, {Borsato}, {Granata},
  {Piotto}, {Lopez}, {Vanderburg}, {Figueira}, {Mortier}, {Nascimbeni},
  {Affer}, {Bonomo}, {Bouchy}, {Buchhave}, {Charbonneau}, {Collier Cameron},
  {Cosentino}, {Dressing}, {Dumusque}, {Fiorenzano}, {Harutyunyan}, {Haywood},
  {Johnson}, {Latham}, {Lopez-Morales}, {Lovis}, {Mayor}, {Micela}, {Molinari},
  {Motalebi}, {Pepe}, {Phillips}, {Pollacco}, {Queloz}, {Rice}, {Sasselov},
  {S{\'e}gransan}, {Sozzetti}, {Udry}, \& {Watson}}]{Malavolta2017}
{Malavolta}, L., {Borsato}, L., {Granata}, V., {et~al.} 2017, \aj, 153, 224

\bibitem[{{Marcy} {et~al.}(2014){Marcy}, {Isaacson}, {Howard}, {Rowe},
  {Jenkins}, {Bryson}, {Latham}, {Howell}, {Gautier}, {Batalha}, {Rogers},
  {Ciardi}, {Fischer}, {Gilliland}, {Kjeldsen}, {Christensen-Dalsgaard},
  {Huber}, {Chaplin}, {Basu}, {Buchhave}, {Quinn}, {Borucki}, {Koch}, {Hunter},
  {Caldwell}, {Van Cleve}, {Kolbl}, {Weiss}, {Petigura}, {Seager}, {Morton},
  {Johnson}, {Ballard}, {Burke}, {Cochran}, {Endl}, {MacQueen}, {Everett},
  {Lissauer}, {Ford}, {Torres}, {Fressin}, {Brown}, {Steffen}, {Charbonneau},
  {Basri}, {Sasselov}, {Winn}, {Sanchis-Ojeda}, {Christiansen}, {Adams},
  {Henze}, {Dupree}, {Fabrycky}, {Fortney}, {Tarter}, {Holman}, {Tenenbaum},
  {Shporer}, {Lucas}, {Welsh}, {Orosz}, {Bedding}, {Campante}, {Davies},
  {Elsworth}, {Handberg}, {Hekker}, {Karoff}, {Kawaler}, {Lund}, {Lundkvist},
  {Metcalfe}, {Miglio}, {Silva Aguirre}, {Stello}, {White}, {Boss}, {Devore},
  {Gould}, {Prsa}, {Agol}, {Barclay}, {Coughlin}, {Brugamyer}, {Mullally},
  {Quintana}, {Still}, {Thompson}, {Morrison}, {Twicken}, {D{\'e}sert},
  {Carter}, {Crepp}, {H{\'e}brard}, {Santerne}, {Moutou}, {Sobeck}, {Hudgins},
  {Haas}, {Robertson}, {Lillo-Box}, \& {Barrado}}]{Marcy2014}
{Marcy}, G.~W., {Isaacson}, H., {Howard}, A.~W., {et~al.} 2014, \apjs, 210, 20

\bibitem[{{Mayor} {et~al.}(2011){Mayor}, {Marmier}, {Lovis}, {Udry},
  {S{\'e}gransan}, {Pepe}, {Benz}, {Bertaux}, {Bouchy}, {Dumusque}, {Lo Curto},
  {Mordasini}, {Queloz}, \& {Santos}}]{Mayor2011}
{Mayor}, M., {Marmier}, M., {Lovis}, C., {et~al.} 2011, ArXiv e-prints
  [\eprint[arXiv]{1109.2497}]

\bibitem[{{Mayor} {et~al.}(2003){Mayor}, {Pepe}, {Queloz}, {Bouchy},
  {Rupprecht}, {Lo Curto}, {Avila}, {Benz}, {Bertaux}, {Bonfils}, {Dall},
  {Dekker}, {Delabre}, {Eckert}, {Fleury}, {Gilliotte}, {Gojak}, {Guzman},
  {Kohler}, {Lizon}, {Longinotti}, {Lovis}, {Megevand}, {Pasquini}, {Reyes},
  {Sivan}, {Sosnowska}, {Soto}, {Udry}, {van Kesteren}, {Weber}, \&
  {Weilenmann}}]{Mayor2003}
{Mayor}, M., {Pepe}, F., {Queloz}, D., {et~al.} 2003, The Messenger, 114, 20

\bibitem[{{McLaughlin}(1924)}]{McLaughlin1924}
{McLaughlin}, D.~B. 1924, \apj, 60, 22

\bibitem[{{Mordasini} {et~al.}(2012){Mordasini}, {Alibert}, {Klahr}, \&
  {Henning}}]{Mordasini2012}
{Mordasini}, C., {Alibert}, Y., {Klahr}, H., \& {Henning}, T. 2012, \aap, 547,
  A111

\bibitem[{{M{\"u}ller} {et~al.}(2013){M{\"u}ller}, {Huber}, {Czesla}, {Wolter},
  \& {Schmitt}}]{Muller2013}
{M{\"u}ller}, H.~M., {Huber}, K.~F., {Czesla}, S., {Wolter}, U., \& {Schmitt},
  J.~H.~M.~M. 2013, \aap, 560, A112

\bibitem[{{Munari} {et~al.}(2014){Munari}, {Henden}, {Frigo}, {Zwitter},
  {Bienaym{\'e}}, {Bland-Hawthorn}, {Boeche}, {Freeman}, {Gibson}, {Gilmore},
  {Grebel}, {Helmi}, {Kordopatis}, {Levine}, {Navarro}, {Parker}, {Reid},
  {Seabroke}, {Siebert}, {Siviero}, {Smith}, {Steinmetz}, {Templeton},
  {Terrell}, {Welch}, {Williams}, \& {Wyse}}]{Munari2014}
{Munari}, U., {Henden}, A., {Frigo}, A., {et~al.} 2014, \aj, 148, 81

\bibitem[{{Nayakshin}(2017)}]{Nayakshin2017}
{Nayakshin}, S. 2017, \pasa, 34, e002

\bibitem[{{Noyes} {et~al.}(1984){Noyes}, {Weiss}, \& {Vaughan}}]{Noyes1984}
{Noyes}, R.~W., {Weiss}, N.~O., \& {Vaughan}, A.~H. 1984, \apj, 287, 769

\bibitem[{{Penev} {et~al.}(2013){Penev}, {Bakos}, {Bayliss}, {Jord{\'a}n},
  {Mohler}, {Zhou}, {Suc}, {Rabus}, {Hartman}, {Mancini}, {B{\'e}ky}, {Csubry},
  {Buchhave}, {Henning}, {Nikolov}, {Cs{\'a}k}, {Brahm}, {Espinoza}, {Conroy},
  {Noyes}, {Sasselov}, {Schmidt}, {Wright}, {Tinney}, {Addison},
  {L{\'a}z{\'a}r}, {Papp}, \& {S{\'a}ri}}]{Penev2013}
{Penev}, K., {Bakos}, G.~{\'A}., {Bayliss}, D., {et~al.} 2013, \aj, 145, 5

\bibitem[{{Pepe} {et~al.}(2002){Pepe}, {Mayor}, {Galland}, {Naef}, {Queloz},
  {Santos}, {Udry}, \& {Burnet}}]{Pepe2002}
{Pepe}, F., {Mayor}, M., {Galland}, F., {et~al.} 2002, \aap, 388, 632

\bibitem[{{Rasio} \& {Ford}(1996)}]{Rasio1996}
{Rasio}, F.~A. \& {Ford}, E.~B. 1996, Science, 274, 954

\bibitem[{{Rodriguez} {et~al.}(2017){Rodriguez}, {Zhou}, {Vanderburg},
  {Eastman}, {Kreidberg}, {Cargile}, {Bieryla}, {Latham}, {Irwin}, {Calkins},
  {Esquerdo}, \& {Mink}}]{Rodriguez2017}
{Rodriguez}, J.~E., {Zhou}, G., {Vanderburg}, A., {et~al.} 2017, ArXiv e-prints
  [\eprint[arXiv]{1701.03807}]

\bibitem[{{Rogers}(2015)}]{Rogers2015}
{Rogers}, L.~A. 2015, \apj, 801, 41

\bibitem[{{Rogers} {et~al.}(2011){Rogers}, {Bodenheimer}, {Lissauer}, \&
  {Seager}}]{Rogers2011}
{Rogers}, L.~A., {Bodenheimer}, P., {Lissauer}, J.~J., \& {Seager}, S. 2011,
  \apj, 738, 59

\bibitem[{{Rossiter}(1924)}]{Rossiter1924}
{Rossiter}, R.~A. 1924, \apj, 60, 15

\bibitem[{{Santerne} {et~al.}(2015){Santerne}, {D{\'{\i}}az}, {Almenara},
  {Bouchy}, {Deleuil}, {Figueira}, {H{\'e}brard}, {Moutou}, {Rodionov}, \&
  {Santos}}]{Santerne2015}
{Santerne}, A., {D{\'{\i}}az}, R.~F., {Almenara}, J.-M., {et~al.} 2015, \mnras,
  451, 2337

\bibitem[{{Santerne} {et~al.}(2012){Santerne}, {Moutou}, {Barros}, {Damiani},
  {D{\'{\i}}az}, {Almenara}, {Bonomo}, {Bouchy}, {Deleuil}, \&
  {H{\'e}brard}}]{Santerne2012b}
{Santerne}, A., {Moutou}, C., {Barros}, S.~C.~C., {et~al.} 2012, \aap, 544, L12

\bibitem[{{Santos} {et~al.}(2015){Santos}, {Adibekyan}, {Mordasini}, {Benz},
  {Delgado-Mena}, {Dorn}, {Buchhave}, {Figueira}, {Mortier}, {Pepe},
  {Santerne}, {Sousa}, \& {Udry}}]{Santos2015}
{Santos}, N.~C., {Adibekyan}, V., {Mordasini}, C., {et~al.} 2015, \aap, 580,
  L13

\bibitem[{{Seager} {et~al.}(2007){Seager}, {Kuchner}, {Hier-Majumder}, \&
  {Militzer}}]{Seager2007}
{Seager}, S., {Kuchner}, M., {Hier-Majumder}, C.~A., \& {Militzer}, B. 2007,
  \apj, 669, 1279

\bibitem[{{Sneden}(1973)}]{Sneden1973}
{Sneden}, C.~A. 1973, PhD thesis, The University of Texas at Austin.

\bibitem[{{Sousa}(2014)}]{Sousa2014}
{Sousa}, S.~G. 2014, {ARES + MOOG: A Practical Overview of an Equivalent Width
  (EW) Method to Derive Stellar Parameters}, ed. E.~{Niemczura}, B.~{Smalley},
  \& W.~{Pych}, 297--310

\bibitem[{{Sousa} {et~al.}(2015){Sousa}, {Santos}, {Adibekyan}, {Delgado-Mena},
  \& {Israelian}}]{Sousa2015}
{Sousa}, S.~G., {Santos}, N.~C., {Adibekyan}, V., {Delgado-Mena}, E., \&
  {Israelian}, G. 2015, \aap, 577, A67

\bibitem[{{Sousa} {et~al.}(2007){Sousa}, {Santos}, {Israelian}, {Mayor}, \&
  {Monteiro}}]{Sousa2007}
{Sousa}, S.~G., {Santos}, N.~C., {Israelian}, G., {Mayor}, M., \& {Monteiro},
  M.~J.~P.~F.~G. 2007, \aap, 469, 783

\bibitem[{{Southworth}(2008)}]{Southworth2008}
{Southworth}, J. 2008, \mnras, 386, 1644

\bibitem[{{Strehl}(1902)}]{strehl1902}
{Strehl}, K. 1902, Astronomische Nachrichten, 158, 89

\bibitem[{{Thiabaud} {et~al.}(2015){Thiabaud}, {Marboeuf}, {Alibert}, {Leya},
  \& {Mezger}}]{Thiabaud2015}
{Thiabaud}, A., {Marboeuf}, U., {Alibert}, Y., {Leya}, I., \& {Mezger}, K.
  2015, \aap, 580, A30

\bibitem[{{Tucci Maia} {et~al.}(2016){Tucci Maia}, {Ram{\'{\i}}rez},
  {Mel{\'e}ndez}, {Bedell}, {Bean}, \& {Asplund}}]{TucciMaia2016}
{Tucci Maia}, M., {Ram{\'{\i}}rez}, I., {Mel{\'e}ndez}, J., {et~al.} 2016,
  \aap, 590, A32

\bibitem[{{Valencia} {et~al.}(2006){Valencia}, {O'Connell}, \&
  {Sasselov}}]{Valencia2006}
{Valencia}, D., {O'Connell}, R.~J., \& {Sasselov}, D. 2006, \icarus, 181, 545

\bibitem[{{Weidenschilling} \& {Marzari}(1996)}]{Weidenschilling1996}
{Weidenschilling}, S.~J. \& {Marzari}, F. 1996, \nat, 384, 619

\bibitem[{{Weiss} \& {Marcy}(2014)}]{Weiss2014}
{Weiss}, L.~M. \& {Marcy}, G.~W. 2014, \apjl, 783, L6

\bibitem[{{Winn} {et~al.}(2010){Winn}, {Fabrycky}, {Albrecht}, \&
  {Johnson}}]{Winn2010}
{Winn}, J.~N., {Fabrycky}, D., {Albrecht}, S., \& {Johnson}, J.~A. 2010, \apjl,
  718, L145

\bibitem[{{Wyttenbach} {et~al.}(2017){Wyttenbach}, {Lovis}, {Ehrenreich},
  {Bourrier}, {Pino}, {Allart}, {Astudillo-Defru}, {Cegla}, {Heng}, {Lavie},
  {Melo}, {Murgas}, {Santerne}, {S{\'e}gransan}, {Udry}, \&
  {Pepe}}]{Wyttenbach2017}
{Wyttenbach}, A., {Lovis}, C., {Ehrenreich}, D., {et~al.} 2017, ArXiv e-prints
  [\eprint[arXiv]{1702.00448}]

\end{thebibliography}

\begin{appendix}

\longtab[1]{
\begin{longtable}{cccccccccc}
\caption{List of the HARPS radial velocity observations of HD~106315  as well as the activity indicators.  \label{rvdata}}\\
\hline
\hline
BJD - 2457000 & RV & $\sigma$ RV & FWHM & $\sigma$ FWHM & BIS & $\sigma$ BIS & S index & $\sigma$ S index & SN50 \\
 days   &    Km$^{-1}$  &  Km$^{-1}$ & Km$^{-1}$ & Km$^{-1}$ & Km$^{-1}$ & Km$^{-1}$ &  &  \\     
  \hline
\endfirsthead
\caption{Continued.} \\
\hline
\hline
BJD - 2457000 & RV & $\sigma$ RV & FWHM & $\sigma$ FWHM & BIS & $\sigma$ BIS & S index & $\sigma$ S index & SN50 \\
 days   &    Km$^{-1}$  &  Km$^{-1}$ & Km$^{-1}$ & Km$^{-1}$ & Km$^{-1}$ & Km$^{-1}$ &  &  \\     
\hline
\endhead
\hline
\endfoot    
781.85390953 & -3.4735 & 0.0046 & 20.5547 & 0.0093 & -0.0044 & 0.0093 & 0.1614 & 0.0008 & 126.4 \\
782.87436772 & -3.4633 & 0.0040 & 20.5316 & 0.0079 & 0.0075 & 0.0079 & 0.1600 & 0.0007 & 90.4 \\
783.81136007 & -3.4673 & 0.0037 & 20.5878 & 0.0075 & -0.0375 & 0.0075 & 0.1615 & 0.0006 & 158.7 \\
783.88784226 & -3.4603 & 0.0037 & 20.5069 & 0.0073 & -0.0153 & 0.0073 & 0.1604 & 0.0006 & 164.4 \\
784.79328906 & -3.4562 & 0.0037 & 20.5735 & 0.0075 & -0.0081 & 0.0075 & 0.1611 & 0.0006 & 157.9 \\
784.87188935 & -3.4625 & 0.0035 & 20.5983 & 0.0071 & 0.0012 & 0.0071 & 0.1595 & 0.0006 & 171.8 \\
785.79689773 & -3.4694 & 0.0038 & 20.5106 & 0.0076 & -0.0099 & 0.0076 & 0.1610 & 0.0006 & 153.6 \\
785.88564972 & -3.4747 & 0.0037 & 20.5331 & 0.0073 & 0.0210 & 0.0073 & 0.1608 & 0.0006 & 164.0 \\
786.82325003 & -3.4638 & 0.0036 & 20.5963 & 0.0072 & -0.0199 & 0.0072 & 0.1598 & 0.0006 & 164.2 \\
786.88285465 & -3.4655 & 0.0034 & 20.5831 & 0.0068 & 0.0331 & 0.0068 & 0.1605 & 0.0005 & 177.4 \\
787.75881115 & -3.4654 & 0.0043 & 20.5866 & 0.0085 & -0.0343 & 0.0085 & 0.1622 & 0.0006 & 134.5 \\
787.85438149 & -3.4750 & 0.0045 & 20.5698 & 0.0090 & 0.0034 & 0.0090 & 0.1636 & 0.0007 & 128.6 \\
789.77546920 & -3.4745 & 0.0040 & 20.5644 & 0.0080 & 0.0010 & 0.0080 & 0.1621 & 0.0006 & 143.5 \\
789.85344400 & -3.4727 & 0.0037 & 20.5702 & 0.0075 & -0.0089 & 0.0075 & 0.1620 & 0.0006 & 157.7 \\
790.80660507 & -3.4737 & 0.0036 & 20.5956 & 0.0072 & 0.0072 & 0.0072 & 0.1606 & 0.0005 & 162.2 \\
790.88801699 & -3.4656 & 0.0036 & 20.5609 & 0.0072 & 0.0105 & 0.0072 & 0.1616 & 0.0006 & 169.2 \\
791.78669995 & -3.4672 & 0.0038 & 20.5483 & 0.0077 & 0.0334 & 0.0077 & 0.1626 & 0.0006 & 149.9 \\
791.87071697 & -3.4699 & 0.0036 & 20.5499 & 0.0072 & -0.0177 & 0.0072 & 0.1622 & 0.0006 & 166.8 \\
792.78409886 & -3.4582 & 0.0034 & 20.5411 & 0.0068 & -0.0064 & 0.0068 & 0.1621 & 0.0005 & 171.8 \\
792.85214066 & -3.4589 & 0.0034 & 20.5711 & 0.0068 & -0.0066 & 0.0068 & 0.1617 & 0.0005 & 176.7 \\
795.74546592 & -3.4534 & 0.0040 & 20.5601 & 0.0080 & -0.0066 & 0.0080 & 0.1602 & 0.0006 & 141.2 \\
795.84097630 & -3.4599 & 0.0038 & 20.5616 & 0.0075 & 0.0137 & 0.0075 & 0.1630 & 0.0006 & 154.9 \\
796.75138530 & -3.4668 & 0.0037 & 20.5390 & 0.0075 & 0.0222 & 0.0075 & 0.1622 & 0.0005 & 154.1 \\
796.85387600 & -3.4591 & 0.0038 & 20.5296 & 0.0076 & 0.0130 & 0.0076 & 0.1657 & 0.0006 & 154.1 \\
797.76222264 & -3.4589 & 0.0038 & 20.5069 & 0.0077 & 0.0430 & 0.0077 & 0.1622 & 0.0006 & 150.0 \\
797.87625292 & -3.4676 & 0.0036 & 20.4928 & 0.0072 & 0.0111 & 0.0072 & 0.1607 & 0.0006 & 167.2 \\
801.78821294 & -3.4649 & 0.0038 & 20.5511 & 0.0075 & 0.0033 & 0.0075 & 0.1614 & 0.0006 & 155.4 \\
801.89787829 & -3.4701 & 0.0051 & 20.5682 & 0.0103 & 0.0190 & 0.0103 & 0.1618 & 0.0009 & 114.4 \\
802.75056942 & -3.4692 & 0.0036 & 20.5470 & 0.0072 & 0.0131 & 0.0072 & 0.1619 & 0.0005 & 160.2 \\
802.86325582 & -3.4621 & 0.0040 & 20.5395 & 0.0080 & 0.0186 & 0.0080 & 0.1616 & 0.0007 & 147.6 \\
803.72859542 & -3.4542 & 0.0043 & 20.5558 & 0.0087 & -0.0240 & 0.0087 & 0.1621 & 0.0007 & 131.0 \\
803.85760258 & -3.4637 & 0.0036 & 20.5883 & 0.0072 & 0.0270 & 0.0072 & 0.1611 & 0.0006 & 166.5 \\
810.72887134 & -3.4693 & 0.0045 & 20.5705 & 0.0090 & 0.0264 & 0.0090 & 0.1597 & 0.0007 & 128.0 \\
810.79508131 & -3.4661 & 0.0036 & 20.5319 & 0.0072 & 0.0132 & 0.0072 & 0.1603 & 0.0006 & 166.5 \\
810.85053677 & -3.4653 & 0.0036 & 20.5459 & 0.0072 & 0.0231 & 0.0072 & 0.1607 & 0.0006 & 166.6 \\
814.72562795 & -3.4644 & 0.0034 & 20.5097 & 0.0068 & -0.0087 & 0.0068 & 0.1595 & 0.0005 & 171.6 \\
814.78416320 & -3.4584 & 0.0035 & 20.5319 & 0.0070 & -0.0152 & 0.0070 & 0.1592 & 0.0006 & 170.2 \\
814.85467687 & -3.4590 & 0.0039 & 20.5821 & 0.0078 & -0.0269 & 0.0078 & 0.1582 & 0.0007 & 152.7 \\
815.70791294 & -3.4626 & 0.0035 & 20.5883 & 0.0070 & 0.0223 & 0.0070 & 0.1593 & 0.0005 & 167.6 \\
815.78306872 & -3.4602 & 0.0036 & 20.6120 & 0.0071 & 0.0030 & 0.0071 & 0.1606 & 0.0006 & 168.1 \\
815.86244959 & -3.4603 & 0.0033 & 20.5888 & 0.0066 & 0.0110 & 0.0066 & 0.1592 & 0.0006 & 189.6 \\
816.87984898 & -3.4603 & 0.0046 & 20.5048 & 0.0091 & 0.0251 & 0.0091 & 0.1617 & 0.0009 & 131.4 \\
817.70931545 & -3.4640 & 0.0039 & 20.5688 & 0.0079 & -0.0138 & 0.0079 & 0.1614 & 0.0006 & 148.6 \\
817.77900779 & -3.4581 & 0.0034 & 20.5740 & 0.0069 & -0.0283 & 0.0069 & 0.1600 & 0.0006 & 173.0 \\
817.84788906 & -3.4636 & 0.0034 & 20.5545 & 0.0069 & -0.0199 & 0.0069 & 0.1600 & 0.0006 & 177.8 \\
820.76823244 & -3.4646 & 0.0036 & 20.5824 & 0.0071 & -0.0200 & 0.0071 & 0.1603 & 0.0006 & 170.9 \\
820.87233534 & -3.4556 & 0.0036 & 20.5720 & 0.0072 & 0.0063 & 0.0072 & 0.1594 & 0.0007 & 171.7 \\
822.70228346 & -3.4661 & 0.0042 & 20.5272 & 0.0083 & 0.0189 & 0.0083 & 0.1605 & 0.0007 & 140.7 \\
822.81967340 & -3.4632 & 0.0038 & 20.5155 & 0.0076 & 0.0195 & 0.0076 & 0.1598 & 0.0006 & 159.9 \\
823.82120368 & -3.4600 & 0.0033 & 20.6031 & 0.0066 & 0.0154 & 0.0066 & 0.1589 & 0.0006 & 187.6 \\
831.68738142 & -3.4571 & 0.0037 & 20.5936 & 0.0074 & -0.0057 & 0.0074 & 0.1612 & 0.0006 & 164.4 \\
834.66790071 & -3.4617 & 0.0044 & 20.5143 & 0.0088 & 0.0019 & 0.0088 & 0.1597 & 0.0007 & 135.0 \\
836.68938313 & -3.4578 & 0.0049 & 20.5885 & 0.0098 & 0.0556 & 0.0098 & 0.1637 & 0.0009 & 121.6 \\
837.65953470 & -3.4653 & 0.0053 & 20.6056 & 0.0105 & 0.0137 & 0.0105 & 0.1631 & 0.0010 & 113.3 \\
838.65577061 & -3.4664 & 0.0038 & 20.5938 & 0.0076 & -0.0095 & 0.0076 & 0.1585 & 0.0007 & 160.1 \\
840.62130041 & -3.4499 & 0.0051 & 20.4929 & 0.0103 & -0.0189 & 0.0103 & 0.1620 & 0.0009 & 114.5 \\
840.71047795 & -3.4653 & 0.0039 & 20.5675 & 0.0078 & -0.0016 & 0.0078 & 0.1612 & 0.0007 & 159.3 \\
843.63219053 & -3.4581 & 0.0043 & 20.5393 & 0.0087 & -0.0127 & 0.0087 & 0.1598 & 0.0007 & 136.4 \\
846.60357079 & -3.4561 & 0.0043 & 20.6207 & 0.0086 & -0.0361 & 0.0086 & 0.1606 & 0.0007 & 138.6 \\
846.76228129 & -3.4662 & 0.0042 & 20.6546 & 0.0084 & -0.0171 & 0.0084 & 0.1625 & 0.0008 & 146.2 \\
847.64546969 & -3.4697 & 0.0052 & 20.5441 & 0.0103 & 0.0014 & 0.0103 & 0.1623 & 0.0010 & 114.1 \\
847.79736299 & -3.4796 & 0.0043 & 20.5497 & 0.0087 & 0.0294 & 0.0087 & 0.1619 & 0.0009 & 143.6 \\
848.59636147 & -3.4671 & 0.0039 & 20.5542 & 0.0079 & 0.0120 & 0.0079 & 0.1608 & 0.0006 & 150.2 \\
848.77503505 & -3.4638 & 0.0041 & 20.6053 & 0.0081 & 0.0378 & 0.0081 & 0.1603 & 0.0008 & 153.1 \\
849.60961949 & -3.4684 & 0.0038 & 20.5916 & 0.0076 & 0.0237 & 0.0076 & 0.1609 & 0.0006 & 157.6 \\
850.63693749 & -3.4672 & 0.0039 & 20.5326 & 0.0077 & 0.0096 & 0.0077 & 0.1620 & 0.0007 & 156.0 \\
850.76055370 & -3.4607 & 0.0038 & 20.5275 & 0.0076 & 0.0115 & 0.0076 & 0.1606 & 0.0007 & 162.9 \\
858.55563723 & -3.4650 & 0.0040 & 20.5953 & 0.0080 & 0.0288 & 0.0080 & 0.1615 & 0.0007 & 148.0 \\
858.75334326 & -3.4580 & 0.0041 & 20.5676 & 0.0083 & 0.0000 & 0.0083 & 0.1608 & 0.0008 & 148.9 \\
859.62681084 & -3.4564 & 0.0036 & 20.5296 & 0.0072 & 0.0138 & 0.0072 & 0.1613 & 0.0006 & 168.4 \\
864.56981350 & -3.4699 & 0.0043 & 20.6523 & 0.0087 & -0.0044 & 0.0087 & 0.1606 & 0.0008 & 138.5 \\
864.70246360 & -3.4688 & 0.0045 & 20.6414 & 0.0090 & 0.0082 & 0.0090 & 0.1598 & 0.0008 & 134.2 \\
865.59395635 & -3.4673 & 0.0040 & 20.5595 & 0.0079 & 0.0679 & 0.0079 & 0.1604 & 0.0007 & 154.7 \\
865.71678168 & -3.4651 & 0.0037 & 20.5752 & 0.0074 & 0.0067 & 0.0074 & 0.1613 & 0.0007 & 170.3 \\
872.63962719 & -3.4560 & 0.0047 & 20.6016 & 0.0094 & -0.0158 & 0.0094 & 0.1641 & 0.0009 & 127.6 \\
872.71917868 & -3.4528 & 0.0038 & 20.6164 & 0.0077 & 0.0046 & 0.0077 & 0.1625 & 0.0008 & 163.8 \\
873.65195481 & -3.4703 & 0.0039 & 20.5755 & 0.0078 & 0.0325 & 0.0078 & 0.1618 & 0.0007 & 157.6 \\
874.63215394 & -3.4482 & 0.0045 & 20.5205 & 0.0090 & -0.0409 & 0.0090 & 0.1613 & 0.0008 & 131.9 \\
875.66309573 & -3.4685 & 0.0074 & 20.5556 & 0.0147 & 0.0487 & 0.0147 & 0.1655 & 0.0019 & 83.4 \\
875.71961729 & -3.4590 & 0.0071 & 20.6324 & 0.0142 & 0.0179 & 0.0142 & 0.1610 & 0.0018 & 86.7 \\
876.57427048 & -3.4572 & 0.0037 & 20.4974 & 0.0074 & -0.0078 & 0.0074 & 0.1620 & 0.0006 & 163.2 \\
876.66817071 & -3.4540 & 0.0036 & 20.5438 & 0.0071 & 0.0070 & 0.0071 & 0.1607 & 0.0006 & 173.3 \\
877.61720792 & -3.4568 & 0.0039 & 20.5600 & 0.0078 & 0.0046 & 0.0078 & 0.1623 & 0.0007 & 156.6 \\
877.71668663 & -3.4534 & 0.0045 & 20.5426 & 0.0090 & 0.0045 & 0.0090 & 0.1629 & 0.0010 & 138.8 \\
821.58826616 & -3.4380 & 0.0134 & 20.5929 & 0.0268 & -0.0470 & 0.0268 & 0.1726 & 0.0042 & 45.6 \\
821.59293130 & -3.4620 & 0.0134 & 20.6711 & 0.0267 & 0.0711 & 0.0267 & 0.1780 & 0.0041 & 45.2 \\
821.59874146 & -3.4452 & 0.0121 & 20.5724 & 0.0242 & -0.0065 & 0.0242 & 0.1709 & 0.0035 & 49.7 \\
821.60311658 & -3.4524 & 0.0115 & 20.6386 & 0.0230 & 0.1648 & 0.0230 & 0.1713 & 0.0031 & 52.1 \\
821.81973748 & -3.4560 & 0.0092 & 20.4562 & 0.0185 & 0.0456 & 0.0185 & 0.1633 & 0.0028 & 66.3 \\
821.82374159 & -3.4536 & 0.0082 & 20.4598 & 0.0163 & 0.0009 & 0.0163 & 0.1682 & 0.0023 & 74.7 \\
821.82815170 & -3.4636 & 0.0085 & 20.5900 & 0.0169 & -0.0270 & 0.0169 & 0.1681 & 0.0025 & 73.4 \\
821.83259682 & -3.4699 & 0.0088 & 20.5595 & 0.0175 & -0.0006 & 0.0175 & 0.1647 & 0.0025 & 70.4 \\
821.83692593 & -3.4771 & 0.0084 & 20.5871 & 0.0169 & 0.0366 & 0.0169 & 0.1643 & 0.0025 & 72.9 \\
821.84157805 & -3.4408 & 0.0097 & 20.6014 & 0.0195 & 0.0992 & 0.0195 & 0.1715 & 0.0031 & 64.2 \\
821.84561815 & -3.4578 & 0.0096 & 20.5806 & 0.0192 & -0.0845 & 0.0192 & 0.1676 & 0.0030 & 64.5 \\
821.85023628 & -3.4564 & 0.0106 & 20.6115 & 0.0212 & 0.0402 & 0.0212 & 0.1689 & 0.0033 & 58.4 \\
821.85456539 & -3.4733 & 0.0091 & 20.4726 & 0.0182 & 0.0560 & 0.0182 & 0.1720 & 0.0028 & 68.0 \\
821.85900950 & -3.4759 & 0.0092 & 20.5772 & 0.0183 & -0.0794 & 0.0183 & 0.1705 & 0.0028 & 68.6 \\
821.86339662 & -3.4431 & 0.0147 & 20.5676 & 0.0294 & 0.0308 & 0.0294 & 0.1682 & 0.0052 & 42.8 \\
821.86776073 & -3.4726 & 0.0091 & 20.6530 & 0.0182 & -0.2816 & 0.0182 & 0.1714 & 0.0030 & 69.0 \\
821.87229784 & -3.4446 & 0.0091 & 20.6435 & 0.0183 & 0.0999 & 0.0183 & 0.1677 & 0.0029 & 69.2 \\
821.87662696 & -3.4799 & 0.0089 & 20.6105 & 0.0177 & -0.1746 & 0.0177 & 0.1658 & 0.0029 & 70.9 \\
821.88115207 & -3.4545 & 0.0092 & 20.6548 & 0.0184 & 0.0352 & 0.0184 & 0.1658 & 0.0030 & 68.3 \\
821.88549318 & -3.4505 & 0.0098 & 20.5941 & 0.0197 & 0.2638 & 0.0197 & 0.1636 & 0.0033 & 64.7 \\
821.88990330 & -3.4662 & 0.0097 & 20.5892 & 0.0193 & -0.0145 & 0.0193 & 0.1631 & 0.0032 & 65.6 \\
821.89431241 & -3.4773 & 0.0099 & 20.5490 & 0.0198 & 0.0607 & 0.0198 & 0.1728 & 0.0034 & 63.8 \\
821.89864152 & -3.4789 & 0.0100 & 20.6267 & 0.0199 & 0.0226 & 0.0199 & 0.1720 & 0.0034 & 63.4 \\
821.90309763 & -3.4736 & 0.0106 & 20.5597 & 0.0212 & 0.0874 & 0.0212 & 0.1709 & 0.0036 & 59.8 \\
821.90784375 & -3.4563 & 0.0118 & 20.7072 & 0.0235 & -0.0580 & 0.0235 & 0.1644 & 0.0047 & 53.8 \\
821.91157085 & -3.4534 & 0.0233 & 21.1203 & 0.0466 & -0.3271 & 0.0466 & 0.1683 & 0.0161 & 29.4 \\
821.91676798 & -3.4571 & 0.0133 & 20.3798 & 0.0265 & -0.2900 & 0.0265 & 0.1682 & 0.0056 & 49.2 \\
842.53900492 & -3.4610 & 0.0121 & 20.4834 & 0.0242 & -0.0195 & 0.0242 & 0.1653 & 0.0037 & 51.4 \\
842.54623887 & -3.4522 & 0.0108 & 20.5691 & 0.0215 & -0.0104 & 0.0215 & 0.1538 & 0.0030 & 56.7 \\
842.55340281 & -3.4710 & 0.0102 & 20.4671 & 0.0204 & 0.0105 & 0.0204 & 0.1551 & 0.0028 & 58.8 \\
842.56070576 & -3.4671 & 0.0105 & 20.6660 & 0.0209 & 1.2301 & 0.0209 & 0.1610 & 0.0029 & 57.7 \\
842.56807870 & -3.4702 & 0.0096 & 20.6130 & 0.0193 & -0.0089 & 0.0193 & 0.1628 & 0.0025 & 62.5 \\
842.57525464 & -3.4645 & 0.0103 & 20.5879 & 0.0206 & 0.0320 & 0.0206 & 0.1578 & 0.0028 & 58.9 \\
842.58268459 & -3.4855 & 0.0093 & 20.5258 & 0.0186 & 0.0337 & 0.0186 & 0.1593 & 0.0024 & 64.5 \\
842.58971053 & -3.4573 & 0.0105 & 20.5380 & 0.0209 & -0.0486 & 0.0209 & 0.1552 & 0.0029 & 58.0 \\
842.59729147 & -3.4703 & 0.0079 & 20.5855 & 0.0158 & 0.0420 & 0.0158 & 0.1581 & 0.0018 & 75.8 \\
842.60424742 & -3.4650 & 0.0090 & 20.5217 & 0.0180 & 0.0101 & 0.0180 & 0.1623 & 0.0023 & 66.5 \\
842.61156136 & -3.4717 & 0.0086 & 20.5593 & 0.0172 & -0.0075 & 0.0172 & 0.1616 & 0.0021 & 69.9 \\
842.61885330 & -3.4690 & 0.0092 & 20.4570 & 0.0184 & 0.0266 & 0.0184 & 0.1573 & 0.0023 & 65.6 \\
842.62615624 & -3.4686 & 0.0092 & 20.5853 & 0.0185 & -0.0010 & 0.0185 & 0.1577 & 0.0024 & 65.0 \\
842.63345918 & -3.4675 & 0.0080 & 20.6186 & 0.0161 & -0.0337 & 0.0161 & 0.1605 & 0.0019 & 74.1 \\
842.64130711 & -3.4700 & 0.0079 & 20.4509 & 0.0157 & 0.0028 & 0.0157 & 0.1583 & 0.0019 & 76.2 \\
842.64840105 & -3.4871 & 0.0077 & 20.5747 & 0.0153 & 0.1201 & 0.0153 & 0.1609 & 0.0018 & 78.3 \\
842.65585499 & -3.4646 & 0.0072 & 20.5599 & 0.0144 & 0.0376 & 0.0144 & 0.1595 & 0.0017 & 83.3 \\
842.66293793 & -3.4617 & 0.0072 & 20.5494 & 0.0143 & -0.0207 & 0.0143 & 0.1633 & 0.0017 & 83.9 \\
842.87929984 & -3.3916 & 0.0336 & 20.3709 & 0.0672 & -0.1926 & 0.0672 & 0.1363 & 0.0413 & 20.7 \\
\hline
\hline
\end{longtable}
}

\begin{table*}
\caption{List of free parameters used in the \texttt{PASTIS} analysis of the light curves, radial velocities, and SED with their associated prior and posterior distribution. }
\begin{center}
\begin{tabular}{lcc}
\hline
\hline
Parameter & Prior & Posterior\\
\hline
\multicolumn{3}{l}{\it Orbital parameters}\\
 &&\\
{\bf HD~106315b}\\
Orbital period $P$ [d] & $\mathcal{N}(9.552; 0.081)$ & 9.55236 $\pm$ 8.8$\times10^{-4}$\\
Epoch of first transit T$_{0}$ [BJD$_{\rm TDB}$ - $2.4\times10^6$] & $\mathcal{N}(57586.5 ; 0.1)$ & 57586.5486 $\pm$ 2.9$\times10^{-3}$\\
Orbital eccentricity $e$ & $\beta(0.867;3.030)$ & 0.078$^{_{+0.110}}_{^{-0.059}}$\\
Argument of periastron $\omega$ [\degree] & $\mathcal{U}(0;360)$ & 245 $^{_{+69}}_{^{-190}}$ \\
Inclination $i$ [\degree] & $\mathcal{S}(70;90)$ & 87.57 $\pm$ 0.28 \\
 &&\\
{\bf HD~106315c}\\
Orbital period $P$ [d] & $\mathcal{N}(21.05; 0.20)$ & 21.05704 $\pm$ 4.3 $\times10^{-4}$\\
Epoch of first transit T$_{0}$ [BJD$_{\rm TDB}$ - $2.4\times10^6$] & $\mathcal{N}(57569.0 ; 0.15)$ & 57569.0173 $\pm$ 1.4$\times10^{-3}$\\
Orbital eccentricity $e$ & $\beta(0.867;3.030)$ & 0.22 $\pm$ 0.16\\
Argument of periastron $\omega$ [\degree] & $\mathcal{U}(0;360)$ & 81 $^{_{+47}}_{^{-32}} $  \\
Inclination $i$ [\degree] & $\mathcal{S}(70;90)$ & 88.66 $^{_{+0.84}}_{^{-0.26}}$ \\
 &&\\
\hline
\multicolumn{3}{l}{\it Planetary parameters}\\
 &&\\
{\bf HD~106315b}\\
Radial velocity amplitude $K$ [\ms] & $\mathcal{U}(0;1000)$ & 3.79 $\pm$ 0.92\\
Planet-to-star radius ratio $k_{r}$ & $\mathcal{U}(0;1)$ & 0.01725$\pm$ 5.1$\times10^{-4}$\\
 &&\\
{\bf HD~106315c}\\
Radial velocity amplitude $K$ [\ms] & $\mathcal{U}(0;1000)$ & 3.11 $\pm$ 0.80\\
Planet-to-star radius ratio $k_{r}$ & $\mathcal{U}(0;1)$ & 0.03073$^{_{+0.00100}}_{^{-0.00069}}$\\
&&\\
\hline
\multicolumn{3}{l}{\it Stellar parameters}\\
 &&\\
Effective temperature \teff\ [K] & $\mathcal{N}(6251;52)$ & 6332 $\pm$ 51\\
Surface gravity \logg\ [g cm$^{-2}$] & $\mathcal{N}(4.1;0.1)$ & 4.258 $\pm$ 0.036\\
Iron abundance [Fe/H] [dex] & $\mathcal{N}(-0.27;0.08)$ & -0.314 $\pm$ 0.078\\
Reddening E(B-V) [mag] & $\mathcal{U}(0;0.1)$ & 0.0049 $^{_{+0.0080}}_{^{-0.0035}}$\\
Systemic radial velocity $\upsilon_{0}$ [\kms] & $\mathcal{U}(-10;5)$ & -3.4648 $\pm$ 6.9 $\times10^{-4}$\\
Distance to Earth $d$ [pc] & $\mathcal{N}(107.3;3.9)$ & 108 $\pm$ 4\\
 &&\\
\hline
\multicolumn{3}{l}{\it Instrumental parameters}\\
 &&\\
HARPS radial velocity jitter [\ms] & $\mathcal{U}(0;1000)$ & 2.89 $\pm$ 0.72\\
SED jitter [mag] & $\mathcal{U}(0;1)$ & 0.078 $\pm$ 0.021\\
\textit{K2} jitter [ppm] & $\mathcal{U}(0;10000)$ & 44 $\pm$ 2\\
\textit{K2} contamination [ppt] & $\mathcal{N}_{\mathcal{U}}(0;5;0;1000)$ & 3.3$^{_{+3.8}}_{^{-2.3}}$\\
\textit{K2} flux normalisation & $\mathcal{U}(0.999;1.001)$ &  1.000001 $\pm$ 3.1 $\times 10^{-6}$\\
LCO jitter [ppt] & $\mathcal{U}(0;100)$ & 1.422 $\pm$ 0.065\\
LCO contamination  & $\mathcal{U}(0;1)$ & 0.10 $^{_{+0.12}}_{^{-0.07}}$\\
LCO flux normalisation & $\mathcal{U}(0.99;1.01)$ & 0.99994  $\pm$ 0.00011\\
GP period [d]  &    $\mathcal{U}(0;8)$   &2.825  $\pm$  0.012\\
GP amplitude  [\ms] &  $\mathcal{U}(0;1)$  & 0.57  $\pm$  0.35 \\
GP $\lambda_1 $  [d]   & $\mathcal{U}(0;1000)$  & 652.5  $\pm$ 340  \\
GP $\lambda_2 $   & $\mathcal{U}(0;1000 )$  &299.0  $\pm$  260 \\
&&\\
\hline
\hline
\end{tabular}
\\
$\mathcal{N}(\mu;\sigma^{2})$ is a normal distribution with mean $\mu$ and width $\sigma^{2}$, $\mathcal{U}(a;b)$ is a uniform distribution between $a$ and $b$, $\mathcal{N}_{\mathcal{U}}(\mu;\sigma^{2},a,b)$ is a normal distribution with mean $\mu$ and width $\sigma^{2}$ multiplied with a uniform distribution between $a$ and $b$, $\mathcal{S}(a,b)$ is a sine distribution between $a$ and $b$, $\beta(a;b)$ is a Beta distribution with parameters $a$ and $b$.
\end{center}
\label{PASTISparams} 
\end{table*}%

\end{appendix}

\end{document}